\documentclass[11pt, a4paper]{article}
\usepackage{t1enc}
\usepackage[latin2]{inputenc}
\usepackage[english]{babel}
\usepackage{fullpage}
\usepackage{caption, subcaption}
\usepackage{graphicx, float}
\usepackage[toc,page]{appendix}
\usepackage{verbatim}
\usepackage{cite}
\usepackage{url}
\usepackage{subfiles}
\usepackage{xcolor}
\usepackage{amsmath} 
\usepackage{amsfonts}
\usepackage{amssymb} 
\usepackage{mathrsfs}
\usepackage{mathtools}
\usepackage{physics}
\usepackage{hyperref}
\usepackage[noabbrev]{cleveref}
\hypersetup{
    unicode=true,          
    pdftoolbar=true,        
    pdfmenubar=true,        
    pdffitwindow=false,     
    pdfstartview={FitH},    
    pdftitle={SEOBNRE CBwaves comparison},    
    pdfauthor={Balázs Kacskovics, Dániel Barta},     
    pdfcreator={Kacskovics Balázs},   
    pdfproducer={Dániel Barta}, 
    pdfkeywords={gravitational waves} {seobnre} {cbwaves} {numerical relativity} {effective-one-body} (post-Newtonian), 
    pdfnewwindow=true,      
    colorlinks=true,       
    linkcolor=red,          
    citecolor=green,        
    filecolor=magenta,      
    urlcolor=cyan           
}

\usepackage{atbegshi}
\AtBeginDocument{\AtBeginShipoutNext{\AtBeginShipoutDiscard}}

\begin{document}

\title{Comparing eccentric waveform models based on post-Newtonian and effective-one-body approaches, over an observationally relevant parameter space}
\author{Bal\'azs Kacskovics, D\'aniel Barta}
\today

\maketitle
\begin{abstract}
    We used two numerical models, namely the \texttt{CBwaves} and \texttt{SEOBNRE} algorithms, based on the post-Newtonian and effective-one-body approaches for binary black holes evolving on eccentric orbits. We performed 20.000 new simulations for non-spinning and 240.000 simulations for aligned-spin configurations on a common grid of parameter values over the parameter space spanned by the mass ratio $q\equiv m_1/m_2\in[0.1,\,1]$, the gravitational mass $m_i \in [10M_\odot,\, 100M_\odot]$ of each component labeled by $i$, the corresponding spin magnitude $S_i \in [0,\,0.6]$ and a constant initial orbital eccentricity $e_{0}$. A detailed investigation was conducted to ascertain whether there was a discrepancy in the waveforms generated by the two codes. This involved an in-depth analysis of the mismatch. Furthermore, an extensive comparison was carried out on the outlier points between the two codes.
\end{abstract}

	\section{Introduction}
	Compact binary coalescences (CBCs) in systems consisting of compact objects (neutron stars and/or black holes) are the main targets for the present generation of ground-based gravitational-wave detectors. The mass ratio and both spins span an 8-dimensional intrinsic parameter space which characterizes the signal associated with an eccentric binary system of spinning compact objects, see \cite{Hannam2014}. Although most analytical waveform models involve the spin parameters due to the fact that black holes can have large spin but often neglect the orbital eccentricity, find more in \cite{Barta2018}. Nevertheless, two potential scenarios can lead to the formation of compact binary systems with non-negligible eccentricity which may fall into the detectable frequency range of terrestrial gravitational-wave observatories. These systems may form in dense stellar environments through dynamical capture with too short time to completely circularize their orbits, or from a dynamic process that increased their initial eccentricity of the system (e.g., Kozai-Lidov oscillations \cite{Lidov1962, Kozai1962}).

	One way to accurately describe the dynamical evolution of CBCs in the inspiral phase is the post-Newtonian approach, where the relative orbital velocity of the components and gravitational potential can be regarded as a small parameter. Thus the motion of the compact objects can be characterized as perturbed Keplerian orbits in a perturbed Schwarzschild metric \cite{BlanchetLRR}. To study the orbital evolution of the binary system, including the precession of spins and eccentricity, we used the computational tool developed in-house, called \texttt{CBwaves} \cite{CBwaves}. The equation of motion is integrated numerically by a fourth-order Runge--Kutta differential equation solver. Then, the field radiation is determined by the simultaneous evaluation of the analytic waveform involving all the contributions up to the 4PN order. As an output of the code, both time and frequency domain waveforms are available. The PN contributions to the acceleration and the radiation field are listed in the appendices of \cite{CBwaves}. Detailed introduction to the center-of-mass PN equation-of-motion and the radiated gravitational radiation of the system can be found in \cite{DDPLA, Kidder, IyerWill, GopaIyer, MoraWill, FBB, FBB2, WillPNSO, WW96, Arun08}.  Furthermore, \texttt{CBwaves} has been expanded with 2.5PN and 3.5PN spin-orbit \cite{Owen, FBB, BoheSOa, BoheSOb} and 4PN \cite{4PN} contribution for the equation-of-motion and derived quantities, and with a more accurate form of the spin precession equation since publishing the original paper \cite{CBwaves}.

	The effective-one-body approach \cite{Buonanno:1998gg, Buonanno:2000ef, Damour2001} is another commonly used analytical model of the general relativistic two-body problem that provides us with a reliable description of dynamics and gravitational waveforms through the inspiral, merger and ringdown stages of the late dynamical evolution of coalescing binary systems \cite{Bohe:2016gbl}. This powerful analytical tool has been significantly improved in the late-inspiral, strong-field, and fast-velocity regimes by the utilization of NR which gives an accurate representation of the merger and ringdown parts in the waveform \cite{Bohe2017, Nagar2017, Damour2014}. The EOBNR models were constructed from the union of EOB and NR models, whose current versions, SEOBNRv3 \cite{Bohe:2016gbl} for nonprecessing binaries and SEOBNRv4 \cite{Babak2017} with the inclusion of BBH precession, are implemented and made them publicly available in the LIGO Scientific Collaboration Algorithm Library (LAL).
	
	In this paper, we explore the accuracy of waveforms computed by \texttt{CBwaves} and \texttt{SEOBNRe}, respectively, and measured by their mismatch over a common parameter space. The parameter space is spaned by the mass-ratios $q\equiv m_1/m_2\in[0.1,\,1]$ with mass $m_i \in [10M_\odot,\, 100M_\odot]$, and spin magnitude $S_i \in [0,\,0.6]$ where $i=1,2$ labels the two black holes. We investigate two cases; one in which the spins are set to be aligned, and one in which they are perpendicular to the orbital plane. Additionally, orbital eccentricity $e$ as an orbital parameter plays a role in the evolution, but due to the fact that it is treated as perturbation in \texttt{SEOBNRe}, its initial value is uniformly $e_{0} = 0.003$.

	\section{CBwaves}
	
	The post-Newtonian approach in the inspiral phase can accurately describe the dynamical evolution of compact binary systems, where the components' relative orbital velocity and gravitational potential can be regarded as small parameters. Thus the motion of the compact objects can be characterized as perturbed Keplerian orbits in a perturbed Schwarzschild metric \cite{BlanchetLRR}.

	To study the orbital evolution of the binary system, including the precession of spins and eccentricity, we used the computational tool, named \texttt{CBwaves} \cite{CBwaves}, developed by the Wigner Gravitational Physics Group. The equation of motion is integrated numerically by a fourth-order Runge--Kutta differential equation solver. Then, the field radiation is determined by the simultaneous evaluation of the analytic waveform involving all the contributions up to the 4PN order. As an output of the code, both time and frequency domain waveforms are available. The PN contributions to the acceleration and the radiation field are listed in the appendices of \cite{CBwaves}. Detailed introduction to the center-of-mass PN equation-of-motion and the radiated gravitational radiation of the system can be found in \cite{DDPLA, Kidder, IyerWill, GopaIyer, MoraWill, FBB, FBB2, WillPNSO, WW96, Arun08}.  Furthermore, \texttt{CBwaves} has been expanded with 2.5PN and 3.5PN spin-orbit \cite{Owen, FBB, BoheSOa, BoheSOb} and 4PN \cite{4PN} contribution for the equation-of-motion and derived quantities, and with a more accurate form of the spin precession equation since the writing of \cite{CBwaves}.
	
	\subsection{Equation of motion in the post-Newtonian expansion}
	As Blanchet and his colleagues have shown in \cite{BlanchetLRR14, 4PN}, the motion of the binary can be derived from its center-of-mass (CM) by imposing vanishing dipole moments for the binary's mass. The integral of the CM, denoted by $ \mathbf{G} $, can be derived from the general-frame Lagrangian, as shown in \cite{BBBFM16, BBBFM17, ABF01, BI03}. The exact form of $ \mathbf{G} $ can be found in the appendix of \cite{4PN}. Resulting from that, one can define the CM frame by solving iteratively the equation $\mathbf{G} = 0$. The CM variables $\mathbf{y}_A$ and $\mathbf{v}_A$ are explicit functions of the relative position $\mathbf{x} = \mathbf{y}_1 - \mathbf{y}_2$ and velocity $\mathbf{v} = \mathbf{v}_1 - \mathbf{v}_2$ in the CM frame given by the iterative solution.  Next, the CM variables were inserted into the equation of motion -- derived from the general-frame Lagrangian -- from which CM acceleration was obtained. Note that the acceleration of the CM will be a function of $\mathbf{n},r,\dot{r}$ and $\mathbf{v}$. Hence, the equation of motion used in \texttt{CBWaves} can be summarized as

		\begin{align}
			\mathbf{a} =& \mathbf{a}_\mathrm{N} + \mathbf{a}_\mathrm{PN} + \mathbf{a}_\mathrm{2PN} + \mathbf{a}_\mathrm{3PN} + \mathbf{a}_\mathrm{4PN} + \mathop{\mathbf{a}}_\mathrm{SO}{}_\mathrm{1.5PN} + \mathop{\mathbf{a}}_\mathrm{SS}{}_\mathrm{2PN} + \mathop{\mathbf{a}}_\mathrm{BT}{}_\mathrm{2.5PN}^\mathrm{RR} + \mathop{\mathbf{a}}_\mathrm{SO}{}_\mathrm{2.5PN} \nonumber\\
 			+& \mathop{\mathbf{a}}_\mathrm{SO}{}_\mathrm{3.5PN} + \mathop{\mathbf{a}}_\mathrm{BT}{}_\mathrm{3.5PN}^\mathrm{RR} + \mathop{\mathbf{a}}_\mathrm{SS}{}_\mathrm{3.5PN}^\mathrm{RR} + \mathop{\mathbf{a}}_\mathrm{SO}{}_\mathrm{3.5PN}^\mathrm{RR}. \label{eq:eom-cbw}
		\end{align}

		The different contributions can be found up to 3.5PN in the appendix of \cite{CBwaves}, and the 4th PN contribution is in \cite{4PN}. Beyond the leading order terms, \cref{eq:eom-cbw} contains extensions for spin-orbit \cite{Owen, FBB, FBB2, BoheSOa, BoheSOb} (SO), spin-spin \cite{Kidder} (SS), and radiation-reaction \cite{IyerWill, Will2005, WW07, Blanchet1993} (RR) effects. 
		\begin{align}
			\mathop{\mathbf{a}}_\mathrm{BT}{}_\mathrm{2.5PN}^\mathrm{RR} =& \frac{8}{5}\eta\frac{G^2 m^2}{c^5 r^3} \left\{ \left[ 3\left( 1 + \beta \right)v^2 + \frac{1}{3}\left( 23 + 6\alpha - 9\beta \right) \frac{Gm}{r} - 5\beta\dot{r}^2 \right]\dot{r}\mathbf{n} \right. \nonumber\\
			&- \left. \left[ \left( 2 + \alpha \right)v^2 + \left( 2 -\alpha \right) \frac{Gm}{r} - 5\beta \dot{r}^2 \right] \right\} \label{eq:radiation-reaction}
		\end{align}
		In equation \ref{eq:radiation-reaction}, one has the freedom of gauge conditions. Depending on our choice of $\alpha$ and $\beta$ gauge parameters, we have a different radiation-reaction term, namely the two known radiation-reaction formulas are the Burke--Thorne and the Damour--Deruelle. In \texttt{CBwaves}, the $\alpha = 4$ and $\beta = 5$ was used in equation \ref{eq:radiation-reaction}. However, if one chooses $\alpha = -1$ and $\beta = 0$ as gauge parameters then the Damour--Deruelle formula is gained, which can be found in \cite{Damour81, Lincoln90}. A detailed description of the gauge parameters can be found in \cite{IyerWill, KaVas22}. The gauge parameters used in $\mathop{\mathbf{a}}_\mathrm{BT}{}_\mathrm{3.5PN}^\mathrm{RR}$ are $\delta_1 \cdots \delta_6$, where in \cite{IyerWill} $\delta_6$ was denoted as $\varepsilon_5$.

	\subsection{Post-Newtonian corrections of the orbital angular momentum, spin and energy}
	The leading order terms of the equation of the motion are derived from a generalized Lagrangian up to the 4th post-Newtonian order. The angular momentum $\mathbf{L}$ and total angular momentum $\mathbf{J} = \mathbf{L} + \mathbf{S}$ can be computed from the Lagrangian. In the absence of radiation reaction terms, these quantities remain conserved up to 2PN order \cite{Kidder}.  Similarly, the total energy is given by the sum of the conserved and radiated energies, $E_\mathrm{tot} = E + \dot{E}$. The conserved energy can be expressed in simple terms, however, the radiated energy $\dot{E}$ is governed by the quadrupole formula, found in \cite{Kidder}. These quantities are evaluated simultaneously with the equation of motion in \texttt{CBwaves}. The angular momentum and total energy expressions are collected in \cite{BlanchetLRR14, 4PN, CBwaves, GopaIyer, Arun08, FBB}. Since the evolution of the spin plays an important role in our analysis, we introduce its evolution equation, for the sake of completeness. In CBwaves, the precession of the spin is governed by
	\begin{equation}
		\mathbf{\dot{S}}_1 = \frac{G}{c^2r^3} \left\{ \Omega _1 \mathbf{L}_N - \mathbf{S}_2 - 3(\mathbf{\hat{n} \cdot S}_2)\mathbf{\hat{n}} + \frac{G^2 \mu m}{c^5 r^2}\left[ \frac{2}{3}(\mathbf{v \cdot S}_2) + 30 \dot{r}(\mathbf{\hat{n}\cdot S}_2) \right]\mathbf{n} \right\}\times \mathbf{S}_1
	\end{equation}
	The above equation contains the standard spin-orbit and spin-spin \cite{Kidder}, the 2PN spin-orbit \cite{FBB2}, the 3PN spin-orbit \cite{BoheSOa}, and the 3.5PN radiative spin--spin \cite{WW96} contributions. Let us note that in CBwaves, we presume the use of covariant spin supplementary condition, $S^{\mu\nu}_A = 0$, where $u_A^\mu$ is the 4-velocity of the center-of-mass world line of the body $A$, with $A = 1,2$. Furthermore, we can find the spin precession correction $\Omega_i$ to the spin-orbit contribution terms in \cite{Kidder, BoheSOa, BoheSOb}. Here, we restrict ourselves to the following form
	\begin{equation}
		\Omega_i = \left[
		\frac{1}{c^2}\alpha_\mathrm{PN} + \frac{1}{c^4}\alpha_\mathrm{2PN} + \frac{1}{c^6}\alpha_\mathrm{3PN}
		+ \mathcal{O}\left(\frac{1}{c^7}\right)\right],
	\end{equation}
	where $\alpha_\mathrm{PN}$, $\alpha_\mathrm{2PN}$ and $\alpha_\mathrm{3PN}$ are called the spin-orbit coefficient by \cite{FBB}.
	\subsection{Radiation field contributions of \texttt{CBwaves}}
		In \texttt{CBwaves}, we determine the field $h_{ij}$ simultaneously with the orbital evolution, as the resulting separation and velocity is essential in its derivation \cite{Kidder}. \texttt{CBwaves} contains the analytic waveform contributions up to 2PN order in harmonic coordinates \cite{Kidder}, valid for general eccentric and spinning sources. The radiation field can be decomposed as the sum of the following contributions

		\begin{eqnarray}\label{eq:04}
			h_{ij} =& \frac{2G\mu}{c^4 D} [Q_{ij} + P^{0.5}Q_{ij} + PQ_{ij}+P^{1.5}Q_{ij}+P^2Q_{ij} + PQ_{ij}^{SO} + P^{1.5}Q_{ij}^{SO} + P^2Q_{ij}^{SO} + PQ_{ij}^{SS}] ~, 
		\end{eqnarray}
	
		where $D$ is the distance to the source from the detector frame, and $\mu = m_1 m_2/(m_1 + m_2)$ is the reduced mass of the binary. Furthermore, $Q_{ij}$ is the quadrupole term, while the rest are higher corrections containing spin-orbit (SO), spin-spin (SS), and the 1.5PN tail corrections \cite{Kidder}. In their fully extended form, these terms can be found in \cite{Kidder, CBwaves}. Similarly to the equation of motion, the radiation field equation can be handily solved by a fourth-order Runge--Kuta differential equation solver to gain the emitted gravitational waves and the orbital evolution with high accuracy.
	
		Looking at a plane wave traveling in direction $\mathbf{\hat{N}}$, which is the unit spatial vector pointing from the center of mass of the source to the observer, the transverse-traceless (TT) part of the radiation field is provided by
		\begin{equation}
			h_{ij}^{TT}= \Lambda_{ij,kl} h_{kl},
		\end{equation}
		where
		\begin{equation}
			\Lambda_{ij,kl}(\mathbf{\hat{N}}) = \mathbf{P}_{ik}\mathbf{P}_{jl} - \frac{1}{2}\mathbf{P}_{ij}\mathbf{P}_{kl}
		\end{equation}
		and
		\begin{equation}
			\mathbf{P}_{ij}(\mathbf{\hat{N}}) = \delta_{ij} - N_i N_j.
		\end{equation}
		
		The radiation field can be defined by choosing an orthonormal triad as in \cite{CBwaves}, where
		\begin{align}
			\mathbf{\hat{N}} &= (\sin \iota \cos \phi, \sin \iota \sin \phi, \cos \iota), \\
			\mathbf{\hat{p}} &= (\cos \iota \cos \phi, \cos \iota \sin \phi, -\sin \iota), \\
			\mathbf{\hat{q}} &= (-\sin \phi, \cos \phi, 0),
		\end{align}
		  and $\iota$ and $\phi$ are the polar angles determining the relative position of the radiation frame with respect to the source frame $(\mathbf{\hat{x}},\mathbf{\hat{y}},\mathbf{\hat{z}})$, see the first figure of \cite{CBwaves}. The polarization states of the gravitation wave can be given, with respect to the orthonormal radiation frame, as
		\begin{equation}
			h_+ = \frac{1}{2}\left(\hat{p}_i\hat{p}_j - \hat{q}_i\hat{q}_j \right)h_{ij}^{TT}, \quad h_\times = \left( \hat{p}_i\hat{q}_j - \hat{q}_i\hat{p}_j \right)h_{ij}^{TT}. \label{eq:hpolar}
		\end{equation}
		The strain produced by the binary system at the detector can be given with the combination
		\begin{equation}
			h(t) = F_+ h_+(t) + F_\times h_\times(t),
		\end{equation}
		where $F_+$ and $F_\times$ are the antenna pattern functions, which can be found in detailed form in \cite{CBwaves}.
		
		After constructing the strain that appears at the detector, we should mention that the radiation field is -- most of the time -- given in terms of spin-weighted spherical harmonics in numerical simulations, just as in \texttt{SEOBNRE} \cite{ZhW17}. Since GW experiments require numerical templates with these modes of the radiation field \cite{Ajith2007}, there is a module implemented in \texttt{CBwaves} to evaluate some of the spin-weighted spherical harmonics as well. The relations, that were applied during the implementation of \texttt{CBWaves} \cite{CBwaves} are the following
		\begin{equation}
			MH_{lm} = \oint \prescript{-2}{}Y^{*}_{lm}(\iota,\phi)(rh_{+} - irh_{\times})\mathrm{d}\Omega,
		\end{equation}
		where -- for instance -- some fo the $\prescript{-2}{}Y^{*}_{lm}$ spherical harmonics read as
		\begin{align}
			\prescript{-2}{}Y_{2 \pm 2} &= \sqrt{\frac{5}{64\pi}} (1 \pm \cos \iota)^2 e^{\pm 2i\phi}, \\
			\prescript{-2}{}Y_{2 \pm 1} &= \sqrt{\frac{5}{16\pi}} \sin \iota (1 \pm \cos \iota) e^{\pm i \phi}, \\
			\prescript{-2}{}Y_{20} &= \sqrt{\frac{5}{32\pi}} \sin^2 \iota.
		\end{align}
		The polarization $h^{(lm)}_+$ and $h^{(lm)}_\times$ are defined as
		\begin{equation}
			rh^{(lm)}_+(t) - irh^{(lm)}_\times(t) = MH_{lm}(t).
		\end{equation}

%
\section{\texttt{SEOBNRE}}
\subsection{Detailed expressions of the conservative EOB Hamiltonian dynamics of \texttt{SEOBNRE}} \label{sec:eob_ham_dyn}

    In this section, we collected the different contributions to the EOB Hamiltonian and related equations that construct the \texttt{SEOBNRE} model \cite{ZhW17}. These equations are distributed within many different papers \cite{BD99, BD2000, DJS2000, BCP07, BPBC07, BBKM08, BPPS09, DN08, DND08, DNH08, DIN09, DN09, DIJS03, DN07}, and for the sake of completeness, they are collected here.

    The idea behind the EOB approach is to reduce the conservative dynamics of the general relativistic two-body problem to a geodesic motion on top of a reduced spacetime corresponding to one body. More precisely, the Mathisson--Pappapetrou--Dixon equation \cite{BRB09} is taken on a deformed Kerr black hole with a metric \cite{BB10} that can be expressed as
    \begin{equation}
        \mathrm{d}s^2 = g_{tt} \mathrm{d}t^2 + g_{rr} \mathrm{d}r^2 + g_{\theta\theta} \mathrm{d}\theta^2 + g_{\phi\phi} \mathrm{d}\phi^2 + 2 g_{t\phi} \mathrm{d}t\mathrm{d}\phi,
    \end{equation}
    where the metric functions can be written as
    \begin{eqnarray*}
        g^{tt} = - \frac{\Lambda_t}{\Delta_t \Sigma}, \quad g^{rr} = \frac{\Delta_r}{\Sigma}, \quad g^{\theta\theta} = \frac{1}{\Sigma}, \\
        g^{\phi\phi} = \frac{1}{\Lambda_t} \left( - \frac{\tilde{\omega}_{fd}^2}{\Delta_t \Sigma} + \frac{\Sigma}{\sin^2 \theta} \right), \qquad g^{t\phi} = - \frac{\tilde{\omega}_{fd}}{\Delta_t \Sigma}
    \end{eqnarray*}
    where
    \begin{align*}
        \Sigma &= r^2 + a^2\cos^2\theta, & \Delta_t &= r^2 \left( A(u) + \frac{a^2u^2}{M^2} \right), & \Delta_r &= \frac{\Delta_t}{D(u)}, \\
        \Lambda_t &= \bar{\omega}^4 - a^2\Delta_t \sin^2 \theta, & \bar{\omega} &\equiv \sqrt{r^2 + a^2}, & \tilde{\omega}_{fd} &= 2Mar + \frac{Ma\eta}{r}\left( \omega^{fd}_1 M^2 + \omega^{fd}_2 a^2 \right).
    \end{align*}
    The equations using the $(t, r, \theta, \phi)$ Boyler--Lindquist coordinates and $\omega^{fd}_1$ and $\omega^{fd}_2$ are set to zero \cite{ZhW17}. Note that, $M$ is the total mass, and $a$ is the Kerr spin parameter, and $M\vec{a} \equiv \vec{\sigma} \equiv \vec{a}_1 m_1 + \vec{a}_2 m_2$, and $u \equiv M/r$ is used, then
    \begin{align}
        A(u) &\equiv 1 - 2u + 2\eta u^3 + \eta \left( \frac{94}{3} - \frac{41}{32}\pi^2 \right) u^4, \\
        D(u) &\equiv \frac{1}{1 + \log\left[ 1 - 6\eta u^2 + 2 \left( 26 - 3\eta \right) \eta u^3 \right]},
    \end{align}
    where $\eta = \frac{m_1*m_2}{M^2}$ is the symmetric mass-ratio. The corresponding Hamiltonian of Mathisson--Pappapetrou--Dixon equations can be written as \cite{BrBu11,TPBB12}
    \begin{equation}
        H = M \sqrt{1 + 2\eta \left( \frac{H_\mathrm{eff}}{M\eta} - 1 \right)} \label{eq:eobham}
    \end{equation}
    where $H_\mathrm{eff} = H_\mathrm{NS} + H_\mathrm{S} + H_\mathrm{SC}$. The three terms -- namely the non-spinning $H_\mathrm{NS}$, the spinning $H_\mathrm{S}$ and the self-spin coupling $H_\mathrm{SC}$ Hamiltonians -- . Based on the Hamiltonian \eqref{eq:eobham}, the equation of motion, with respect to the conservative part, takes the form of
    \begin{equation}
        \dot{\vec{r}} = \frac{\partial H}{\partial \vec{\tilde{p}}}, \qquad \dot{\vec{\tilde{p}}} = - \frac{\partial H}{\partial \vec{r}}. \label{eq:eob-eom}
    \end{equation}

\subsection{Radiation field equations and their spherical modes used in the eccentric waveform}
    For a similar reason as in the case of \texttt{CBWaves}, in the EOBNR framework, spin-weighted $-2$ spherical harmonic modes of the gravitational radiation field were considered during the creation of the model. These modes are extensively used in numerical relativity \cite{BCHL11} as well. In the mainstream code, \texttt{SEOBNRv1}, from the modes $l \in \{ 2,3,4,5,6,7,8\}$ and $m \in [-l,l]$ only the positive ones were implemented. Note that, the negative modes available via $h_{lm} = (-1)^l h^*_{l,-m}$ relation \cite{PBK11}. However, in \texttt{ SEOBNRE} only $(2,2)$ mode was used. Basically in EOBNR models, the radiation field equations are decomposed into two parts, a quasicircular and an eccentric part, following the strategy introduced in \cite{HKSz17}. We have to mention at this point, that the eccentricity in \texttt{SEOBNRE} was treated as a perturbation by assuming it as a relatively small parameter. In the EOBNR framework, the quasicircular part of the radiation field is divided into two segments, namely the inspiral-plunge and post-merger phase. The post-merger phase is described with the quasinormal modes of a Kerr black hole. While the inspiral-plunge stage can be described in the factorized form as \cite{PBK11}
    \begin{align}
        h^{(C)}_{lm} &= h^{(N,\epsilon)}_{lm} \hat{S}^{(\epsilon)}_\mathrm{eff} T_{lm} e^{i\delta_{lm}}(\rho_{lm})^l N_{lm}, \label{eq:hClm} \\
        h^{(N,\epsilon)}_{lm} &= \frac{M\eta}{D} n^{(\epsilon)}_{lm} c_{l+\epsilon} V^l_\Phi Y^{l-\epsilon,-m}\left( \frac{\pi}{2}, \Phi \right), \label{eq:hNepslm}
    \end{align}
    where $D$ is the distance of the source from the detector frame, $\Phi$ is the orbital phase, and $Y^{lm}(\Theta,\Phi)$ are the scalar spherical harmonics. In particular for the $(2,2)$ mode, the values of $\epsilon$ and $V_\Phi^2$ are set to be \cite{ZhW17}
    \begin{equation}
        \epsilon = 0, \qquad V^2_\Phi = v^2_\Phi, \qquad \vec{v}_\Phi = \vec{v}_p - \vec{n} \left( \vec{v}_p \cdot \vec{n} \right), \qquad \vec{v}_p \equiv \dot{\vec{r}} = \frac{\partial H}{\partial \vec{\tilde{p}}}.
    \end{equation}
    Then in eq. \eqref{eq:hClm}, $T_{lm}$ re-summates the leading-order contributions of the tail effect, while the function $\hat{S}_\mathrm{eff}^{(\epsilon)}$ is an effective source term, which in the circular-motion limit contains a pole at the EOB light ring. With the above conditions for the $(2,2)$ modes, they can be given as \cite{PBK11,TPBB12,ZhW17}
    \begin{equation}
        T_{lm} = \frac{\Gamma(l + 1 - 2im\Omega H)}{\Gamma(l + 1)}e^{\pi m \Omega H + 2im\Omega H\ln(2m\Omega r_0)} \qquad \hat{S}^{(0)}_\mathrm{eff} = \frac{H_\mathrm{eff}}{M\eta},
    \end{equation}
    where
    \begin{equation}
        \Omega \equiv v_\Phi^3, \qquad r_0 \equiv \frac{2 (m_1 + m_2)}{\sqrt{e}}.
    \end{equation}
    The $\exp(i\delta_{lm})$ in eq. \eqref{eq:hClm} is a phase correction due to subleading-order corrections, and the $(\delta_{lm})^l$ component contains the remaining PN terms, and $N_{lm}$ is the non-quasicircular (NQC) corrections, that model the deviations from the quasicircular motion. As it can be found in \texttt{SEOBNRE}\cite{ZhW17}, the extended form of these components of eq. \eqref{eq:hClm} are
    \begin{align}
        \delta_{22} =& \frac{7}{3}\bar{v}^3 + \left( \frac{428}{105}\pi - \frac{4}{3}a \right)\bar{v}^6 + \left( \frac{1712}{315}\pi^2 - \frac{2203}{81} \right) \bar{v}^9 - 24\eta v_\Phi^5 + \frac{20}{63}a v_\Phi^8 \\
        \rho_{22} =& 1 + \left( \frac{55}{84}\eta - \frac{43}{42} \right) v_\Phi^2 - \frac{2}{3} \left[ \xi_S \left( 1 - \eta\right) + \xi_A \sqrt{1 - 4\eta} \right] v_\Phi^2 \nonumber \\
        &+ \left( \frac{a^2}{2} - \frac{20555}{10584} - \frac{33025}{21168}\eta + \frac{19583}{42336}\eta^2 \right) v_\Phi^4 + \left( \frac{1556919113}{122245200} + \frac{89}{252}a^2 \right. \nonumber \\
        &- \left. \frac{48993925}{9779616}\eta - \frac{6292061}{3259872}\eta^2 + \frac{10620745}{39118464}\eta^3 + \frac{41}{192}\pi^2 \eta - \frac{428}{105} \log^\mathrm{E}_2(v_\Phi^2) \right) v_\phi^6 \nonumber \\
        &+ \left( \frac{18733}{15876}a + \frac{1}{3}a^3 \right)v_\phi^7 + \left[ \frac{18353}{21168}a^2 - \frac{1}{8}a^4 - (5.6 + 117.6\eta)\eta - \frac{387216563023}{160190110080} \right. \nonumber \\
        &+ \left. \frac{9202}{2205}\log_2^\mathrm{E}(v_\Phi^2) \right] v_\Phi^8 - \left( \frac{16094530514677}{533967033600} - \frac{439877}{55566}\log_2^\mathrm{E}(v_\Phi^2) \right)v_\Phi^10 \\
        N_{lm} =& \left[ 1 + \frac{\tilde{p}_r^2}{(r\Omega)^2} \left( a_1^{h_{lm}} + \frac{a_2^{h_{lm}}}{r} + \frac{a_3^{h_{lm}} + a_{3S}^{h_{lm}}}{r^{3/2}} + \frac{a_4^{h_{lm}}}{r^2} + \frac{a_5^{h_{lm}}}{r^{5/2}} \right) \right] \nonumber \\
        &\times \exp\left[ i\left( \frac{\tilde{p}_r }{r\Omega} b_1^{h_{lm}} + \frac{\tilde{p}_r^3}{r\Omega} \left( b_2^{h_{lm}} + \frac{b_3^{h_{lm}}}{r^{1/2}} + \frac{b_4^{h_{lm}}}{r} \right) \right) \right]
    \end{align}
    where $\log^\mathrm{E}_m(v_\Phi^2) \equiv \mathrm{eulerlog}_m(v_\Phi^2) = \gamma_\mathrm{E} + \log 2 + \log m + 1/2 \log v_\Phi^2$, which is used in \texttt{SEOBNRE} in the following form $\log_m^\mathrm{E}(v_\Phi^2) \equiv \gamma_E + \ln(2mv_\Phi)$ with $\gamma_\mathrm{E}$ being the Euler constant, which exact value can be found in \cite{ZhW16,ZhW17}. Furthermore, $a_i^{h_{lm}}$, with $i \in [1,5]$, and $a_{3S}^{h_{lm}}$, and $b_i^{h_{lm}}$ with $i \in [1,4]$ are respectively the real NQC amplitude and phase coefficients. The NQC coefficients can be determined numerically for specific cases in $a$ and $\eta$ by the procedure recommended in \cite{TPBB12}. Then the exact values of $a$ and $\eta$ are interpolated in \texttt{SEOBNRE} \cite{ZhW17}, then the conditions (21)--(25) from \cite{TPBB12} are solved.

    The remaining two functions in eq. \eqref{eq:hNepslm} are can be writen as \cite{ZhW17,TPBB12,PBK11}
    \begin{equation}
        n^{(0)}_{lm} = (im)^l \frac{8\pi}{(2l + 1)!!} \sqrt{\frac{(l + 1)(l + 2)}{l(l - 1)}} \qquad c_l = \left( \frac{m_2}{m_1 + m_2} \right)^{l-1} + (-1)^l \left( \frac{m_1}{m_1 + m_2} \right)
    \end{equation}

    For the eccentric part, in the radiation field terms up to the second post-Newtonian order are considered. As such, the PN radiation field $h^{ij}$ can be expressed as
    \begin{equation}
        h^{ij} = 2\eta \left( Q^{ij} + P^{1/2}Q^{ij} + PQ^{ij} + P^{3/2}Q^{ij} + P^{3/2}Q^{ij}_\mathrm{tail} \right),
    \end{equation}
    where the exact form of each contribution can be found in \cite{WW96, ZhW17, CBwaves}, and the tail component can be expressed as \cite{WW96}
    \begin{equation}
        P^{3/2} Q^{ij}_\mathrm{tail} = 4 v_p^5 \left[ \pi \left( \lambda^i \lambda^j - n^j n^i \right) + 6\ln v_p \left( \lambda^i n^j + n^i \lambda^j \right) \right], \label{eq:rfcont_3.5tail}
    \end{equation}
    where
    \begin{equation}
        \vec{\lambda} = \frac{\vec{v}_p - (\vec{v}_p \cdot \vec{n}) \vec{n}}{\left| \vec{v}_p - (\vec{v}_p \cdot \vec{n}) \vec{n} \right|}.
    \end{equation}
    The definition of the polarization of $h^{ij}$ is the same as eq. \eqref{eq:hpolar}, however note that
    \begin{align}
        \epsilon^{+}_{ij} &= \frac{1}{2}\left( \hat{p}_i \hat{p}_j - \hat{q}_i \hat{q}_j \right) \\
        \epsilon^{\times}_{ij} &= \frac{1}{2}\left( \hat{p}_i \hat{q}_j + \hat{q}_i \hat{p}_j \right)
    \end{align}
    are the polarisation vectors, and $h = h_+ - ih_\times$. Then the spherical modes of the radiation field can be calculated by
    \begin{equation}
        h_{l,m} = \int h^{-2} Y^*_{l,m} d\Omega.
    \end{equation}
    Based on the above integral, the $(2,2)$ mode of the gravitational radiation can be expressed as
    \begin{align}
        h_{22} =& 2\eta \left[ \Theta_ij \left( Q^{ij} + P_0 Q^{ij} + P^{3/2}Q^{ij}_\mathrm{tail} \right) + P_n \Theta_{ij} \left( P^{1/2}_{n}Q^{ij} + P^{1/2}_{n}Q^{ij} \right) + P_v \Theta_{ij} \left( P^{1/2}_{v}Q^{ij} + P^{1/2}_{v}Q^{ij} \right) \right. \nonumber \\
        &+ P_{nn}\Theta_{ij}P_{nn}Q^{ij} + P_{nv}\Theta_{ij}P_{nv}Q^{ij} + P_{vv}\Theta_{ij}P_{vv}Q^{ij} + P_{nnn}\Theta_{ij}P^{3/2}_{nnn}Q^{ij} + P_{nnv}\Theta_{ij}P^{3/2}_{nnv}Q^{ij} \nonumber \\
        &+ \left. P_{nvv}\Theta_{ij}P^{3/2}_{nvv}Q^{ij} + P_{vvv}\Theta_{ij}P^{3/2}_{vvv}Q^{ij} \right]. \label{eq:h22mode}
    \end{align}
    It is assumed by the authors of \texttt{SEOBNRE} \cite{ZhW17}, that in eq. \eqref{eq:h22mode} a quasicircular part is included, which corresponds to $h_{22}|_{\dot{r}=0}$ and the left eccentric part. To check, wether $h_{22}|_{\dot{r}=0}$ is consistent with eq. $(9.3)$ of \cite{BFIS08, BFIS12}, is very straightforward, so the eccentric correction can be defined as
    \begin{equation}
        h_{22}^{(\mathrm{PNE})} = h_{22} - h_{22}|_{\dot{r}=0},
    \end{equation}
    where $h_{22}$ is same as eq. \eqref{eq:h22mode}. In short, the inspiral-plunge waveform in \texttt{SEOBNRE} is
    \begin{equation}
        h^\mathrm{inspiral-plunge}_{22} = h_{22}^{(\mathrm{C})} + h_{22}^{(\mathrm{PNE})} \label{eq:seobnre-wf}
    \end{equation}
    where $h_{22}^{(\mathrm{C})}$ is given by eq. \eqref{eq:hClm}.

\subsection{The radiation-reaction force for the EOB dynamics and the post-Newtonian energy flux for an eccentric binary}
    In \cref{sec:eob_ham_dyn}, the conservative part of the EOB dynamics was discussed, culminating in the \cref{eq:eob-eom} of motion. However, it is only one-half of the EOB dynamics, the left side of eq. \eqref{eq:eob-eom} is also related to the radiation-reaction force $\vec{\mathcal{F}}$. Taking into consideration the radiation-reaction force, the \cref{eq:eob-eom} becomes
    \begin{equation}
        \dot{\vec{r}} = \frac{\partial H}{\partial \vec{\tilde{p}}}, \qquad \dot{\vec{\tilde{p}}} = - \frac{\partial H}{\partial \vec{r}} + \vec{\mathcal{F}}. \label{eq:eob-rreom}
    \end{equation}
    The radiation-reaction force related to the energy flux $\dot{E}$ of the gravitational radiation via \cite{TPBB12}
    \begin{equation}
        \vec{\mathcal{F}} = \frac{1}{M\eta \omega_\Phi \left| \vec{r} \times \vec{p} \right|} \dot{E} \vec{\tilde{p}},
    \end{equation}
    where $\omega_\Phi$ is the orbital frequency, which can be written as
    \begin{equation}
        \omega_\Phi = \frac{\left| \vec{r} \times \dot{\vec{r}} \right|}{r^2}.
    \end{equation}
    Calling attention to the fact that the sign of $\dot{E}$ is negative since $E$ is the energy of the binary, and it decreases with the radiated gravitational waves, so $\dot{E} < 0$ at all times. This is why sometimes, we are referred to it as dissipation. \texttt{SEOBNRv1} and similarly to it \texttt{SEOBNRE} code treats quasicircular case without precession or $\left| \vec{r} \times \vec{p} \right| \approx \tilde{p}_\Phi$, which reduces the radiation-reaction force into
    \begin{equation}
        \vec{\mathcal{F}} = \frac{1}{M\eta \omega_\Phi} \dot{E} \frac{\vec{\tilde{p}}}{\tilde{p}_\Phi}
    \end{equation}
    
    In the \texttt{SEOBNR} model, the energy-flux $\dot{E}$ related to the waveform of the gravitational radiation through \cite{PBF11}
    \begin{equation}
        -\dot{E} = \frac{1}{16\pi} \sum \limits_l \sum \limits_{m = -l}^l \left| \dot{h}_{lm} \right|^2. \label{eq:ener-flux-seo}
    \end{equation}
    The more it is assumed that $h_{lm}$ is dependent on time, the more it is similar to a harmonic oscillation. Accordingly $\dot{h}_{lm} \approx m\Omega h_{lm}$ with $\Omega$ being the orbital frequency. What's more, the relation between the radiation field and the energy flux becomes
    \begin{equation}
        -\dot{E} = \frac{1}{16\pi} \sum \limits_l \sum \limits_{m = -l}^l (m\Omega)^2 \left| h_{lm} \right|^2 = \frac{1}{8\pi} \sum \limits_l \sum \limits_{m = 1}^l (m\Omega)^2 \left| h_{lm} \right|^2 \label{eq:ener-flux-red}
    \end{equation}
    Just like in EOBNR models, the equation system implemented in \texttt{SEOBNRE} is valid strictly for spin-aligned binary blackholes, where the spins of the two components of the binary are perpendicular to the orbital plane. In these kinds of systems, no precessions are involved during orbital evolution, and more importantly, there is a plane reflection symmetry to the orbital plane. Thanks to this symmetry of the corresponding spacetime and the spin-weighted spherical harmonic functions, one has \cite{BGH08}
    \begin{equation}
        h(t, \pi - \theta, \phi) = h^*(t, \theta, \phi), \qquad \prescript{-2}{}Y_{l,-m}(\pi - \theta, \phi) = \prescript{-2}{}Y^*_{lm}(\theta, \phi),
    \end{equation}
    where $\theta$ and $\phi$ are the spherical coordinates with respect to the gravitational wave source. As a consequence of the above equations, $h_{lm} = (-1)^lh^*_{l,-m}$ \cite{TPBB12}, which was taken into consideration in the second equality of the energy flux in eq. \eqref{eq:ener-flux-red}. With this, the so-called ''memory modes'' of the radiation field $h_{l0}$ were neglected in the SEOBNR model \cite{ZhW16, Nichols17}.
    
    In their \texttt{SEOBNRE} model \cite{ZhW17}, they followed the steps of the \texttt{SEOBNRv1} to construct the radiation-reaction force, except they used the \texttt{SEOBNRE} waveform explained in eq. \eqref{eq:seobnre-wf}. Beyond the method introduced above to calculate the energy flux, there are other ways to do it. A widely used method is to derive the energy flux by utilizing the results of the conservative EOBNR Hamiltonian dynamics within the post-Newtonian framework. This method was used in the ax model \cite{HKA17}. The energy flux can treated the same way as the radiation was previously. It can be divided into a quasicircular, an elliptic, and a noncircular correction part. In the end, the overall energy flux can be written as
    \begin{equation}
        \dot{E} = \dot{E}_{{\mathrm{C}}} + \dot{E}_\mathrm{Elip} - \dot{E}_{\mathrm{Elip}, \dot{r} = 0},
    \end{equation}
    where the exact form of the contributions in the above equation can be found in \cite{ZhW17}.

	\section{Numerical Results}
		In this work, we have compared the gravitational waveforms produced by the two aforementioned codes. This has been done using an identical initial parameter space for the two codes. Since the two codes are based on two different extensions of the Newtonian theory, we had to be careful with our choice of initial parameters. During our simulations, we choose the initial separation to be $30~\mathrm{M}$, where $\mathrm{M} = m_1 + m_2$ is the total mass of the binary. However, \texttt{SEOBNRE} code takes the initial orbital frequency as input and computes the separation from it. To match the initial parameter, we calculated the initial orbital frequency by using
		\begin{equation}
			f_\mathrm{init} = \frac{c^3}{\pi G (m_1 + m_2) M_\odot \sqrt{\mathfrak{r}_0^3}},
		\end{equation}
		where $\mathfrak{r}_0 = 30$ is the dimensionless initial separation, and $M_\odot$ is the solar mass. Also, it has to be mentioned that \texttt{SEOBNRE} is valid only within the Kerr-bound -- as long as $\norm{\mathbf{S}} < 1$ --, and the dimensionless spin parameter falls within the $\chi \in [-1, 0.6]$ range. It should also be noted that the eccentricity in SEOBNRE was treated as a small perturbation and, in line with \cite{ZhW17},  we set it to $e_0 = 0.003$. In order to constrain our simulations within the limits of the parameter space in which the \texttt{SEOBNRE}is valid, we picked six values of the dimensionless spin parameter from $(0.1, 0.6)$. As stated above, only spin-aligned configurations can be calculated by \texttt{SEOBNRE}, while this is not true for\texttt{CBwaves}. Therefore, with \texttt{CBwaves}, computations were performed for both aligned and non-aligned spin configurations. For the sake of completeness, a non-spinning computation was made using each code.

		In their work, the authors \texttt{CBwaves} code, the authors \cite{CBwaves} developed the CBwaves code, enabling the interchangeability of both the equation of motion and the radiation field contributions to the post-Newtonian expansion. Hence, we choose our contributions to match \texttt{SEOBNRE} as much as possible. The used contributions are
		\begin{align}
			\mathbf{a} =& \mathbf{a}_\mathrm{N} + \mathbf{a}_\mathrm{PN} + \mathbf{a}_\mathrm{2PN} + \mathbf{a}_\mathrm{3PN} + \mathop{\mathbf{a}}_\mathrm{SO}{}_\mathrm{1.5PN} + \mathop{\mathbf{a}}_\mathrm{SS}{}_\mathrm{2PN} + \mathop{\mathbf{a}}_\mathrm{BT}{}_\mathrm{2.5PN}^\mathrm{RR} + \mathop{\mathbf{a}}_\mathrm{SO}{}_\mathrm{2.5PN} \nonumber\\
 			+& \mathop{\mathbf{a}}_\mathrm{SO}{}_\mathrm{3.5PN} + \mathop{\mathbf{a}}_\mathrm{SS}{}_\mathrm{3.5PN}^\mathrm{RR} + \mathop{\mathbf{a}}_\mathrm{SO}{}_\mathrm{3.5PN}^\mathrm{RR}, \label{eq:eom-cbw-used} \\
			 h_{ij} =& \frac{2G\mu}{c^4 D} [Q_{ij} + P^{0.5}Q_{ij} + PQ_{ij}+P^{1.5}Q_{ij} + PQ_{ij}^{SO}] ~. 
		\end{align}

		To achieve our primary goal, to map the mismatch between \texttt{SEOBNRE} and \texttt{CBwaves} over a parameter space spanned by $m_1$ and $m_2$ with high accuracy, we had to perform an enormous amount of computer simulation. We decided to investigate the range $(10~\mathrm{M}_\odot, 100~\mathrm{M}_\odot)$ of gravitational mass in both $m_1$ and $m_2$. We took $100$ points between the two values for both components and since we had two codes and thirteen spin configurations, which led to $260000$ computations. To complete this task in a reasonable time, we used a HTCondor cluster to carry out multiple evaluations in parallel.

\begin{figure}[!h]
			\begin{center}
				\includegraphics[width = 0.8 \textwidth]{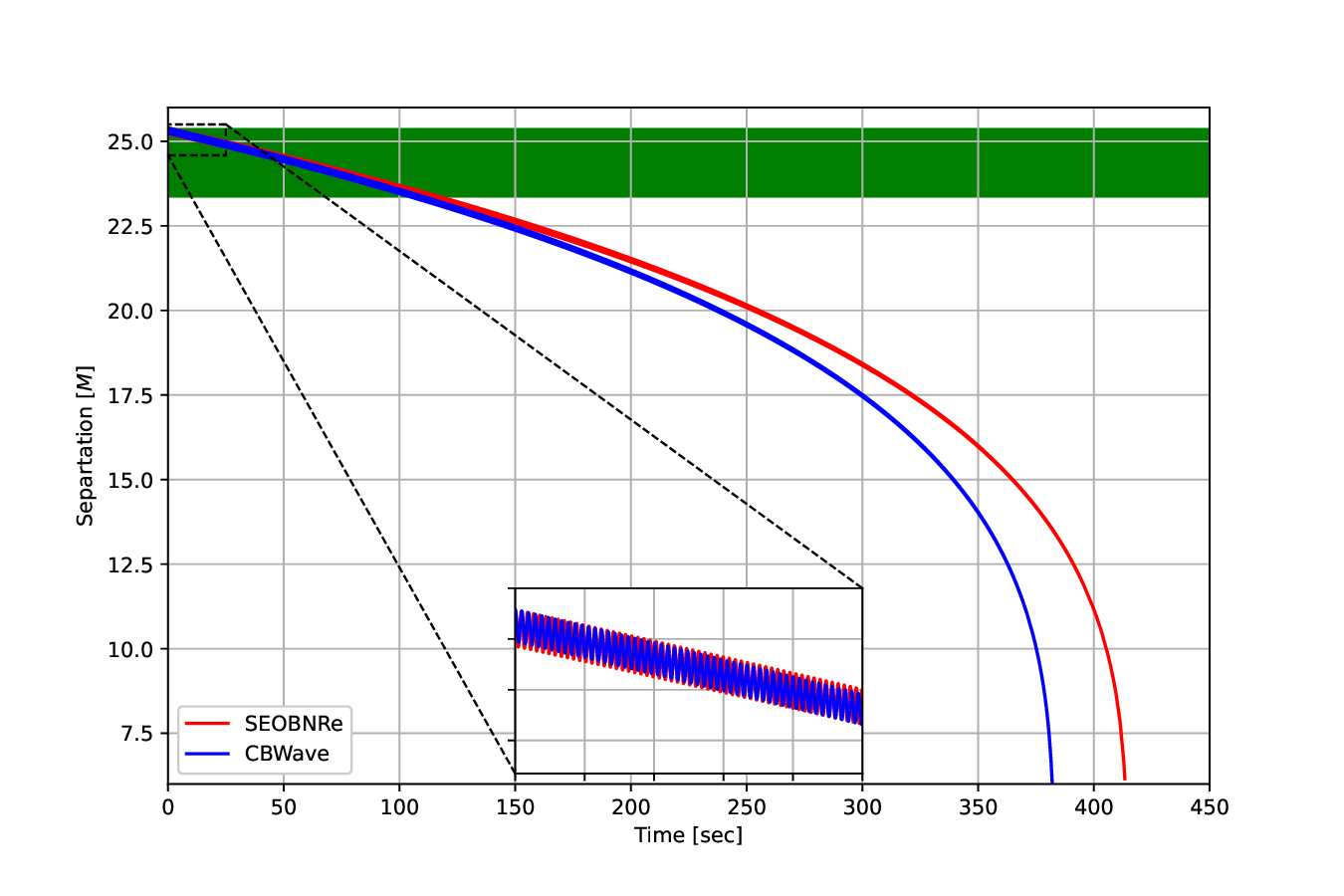}
				\caption{The figure displays the evolution of the separation of the two compact bodies at $\chi_1 = 0.6\dot{2}$. The green band represents the range from where the sample was taken for the mismatch calculation. We also verify here whether the orbital evolution computed by each code starts at the same initial separation, which we enlarged in an inset figure for the first couple of seconds.} \label{fig:separation-diff-s1-06}
			\end{center}
		\end{figure}

\begin{figure}[!h]
			\begin{center}
				\includegraphics[width = 0.8 \textwidth]{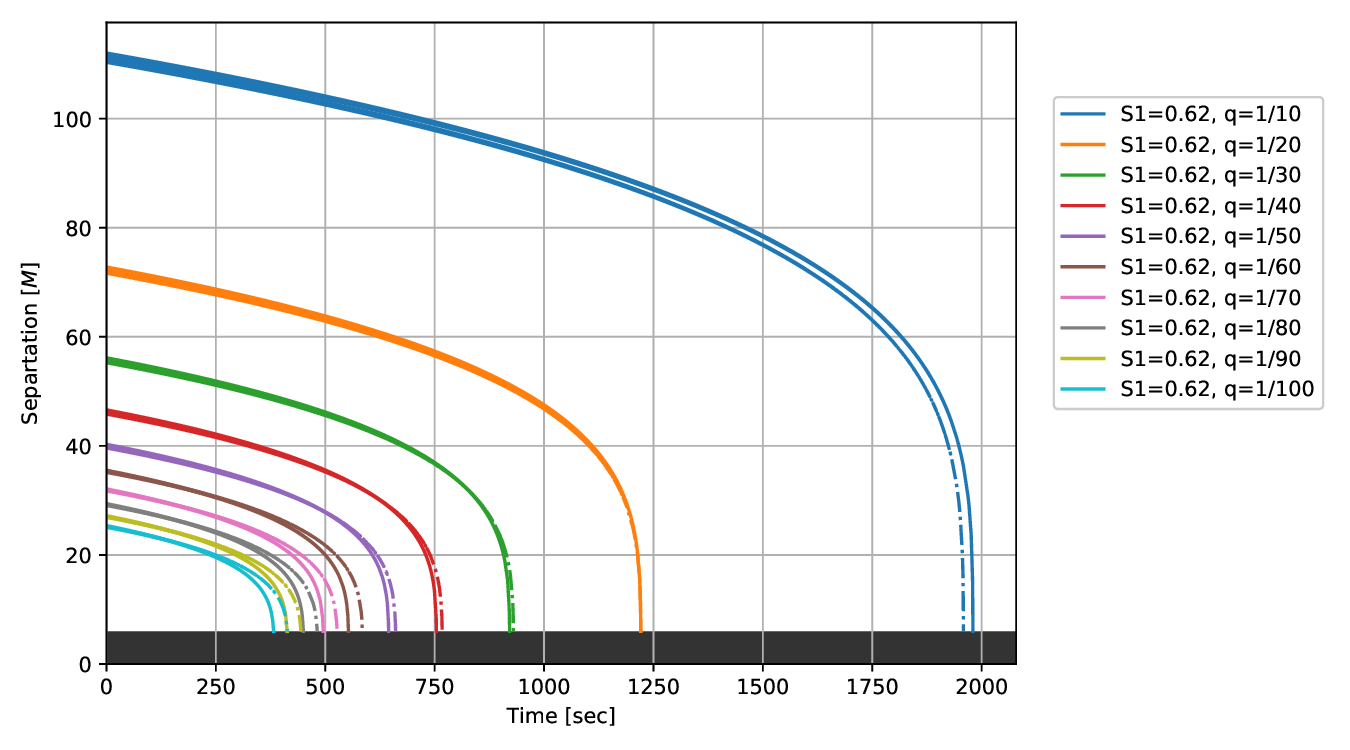}
				\caption{The evolution of the separation is displayed at different mass ratios with the same initial orbital frequency and dimensionless spin parameter. Here, the dark grey area represents the near zone, where the separation is less than $6~\mathrm{M}$. Separations calculated with \texttt{CBwave} are represented with continuous lines, while dashed lines represent \texttt{SEOBNRE}.} \label{fig:separation-diffs-s1-06}
			\end{center}
		\end{figure}

		To calculate the mismatch, we had to find the time interval where the waveforms generated by both programs were the most similar.  Since \texttt{SEOBNRE} evaluates the entire coalescence, hence orbits beyond the last stable orbits (LSO) (orbits with a separation smaller than $6~\mathrm{M}$), we have to cut those orbits out of the analyzed data. Then, we have chosen a part of the orbital evolution of the binary, in which the development of the separations, calculated by both codes, was the closest. Fig. \ref{fig:separation-diff-s1-06} shows how we chose the section of the times series where the mismatch was calculated on the separation for a configuration selected by us. Note that, in Fig. \ref{fig:separation-diff-s1-06} and \ref{fig:separation-diffs-s1-06}, the initial separation was calculated from the initial orbital frequency, which was $f_0 = 5~\mathrm{Hz}$. The mass ratio of the binary was $1:100$ in Fig. \ref{fig:separation-diff-s1-06}, while for Fig. \ref{fig:separation-diffs-s1-06} $m_1$ spanned between 10 and 100 solar mass, and $m_2$ was set to one solar mass. Also, the dimensionless spin parameter of the primary components was set to $0.62$ to see how \texttt{SEOBNRE} performs a bit above its limitations.

		From Fig.  \ref{fig:separation-diff-s1-06}, it is apparent that there is a significant difference in the orbital evolution time of the two components until it reaches the $6~\mathrm{M}$ separation limit. Roughly at $150~\mathrm{seconds}$, the gap between the two curves gets progressively wider and peaks at around $31~\mathrm{seconds}$. The magnified area in the figure shows that both codes started at the same separation. However, to achieve it, we have to use a slightly smaller orbital frequency to calculate the separation for \texttt{CBwaves}. This implies that some rounding differences occur between the results of the two codes during the evaluation of the initial conditions from the input parameters. After seeing the difference in the orbital evolution for an individual configuration, we deemed it necessary to investigate configurations with different mass ratios, as shown in Fig. \ref{fig:separation-diffs-s1-06}.

		As the mass ratio decreases, the difference in time required to reach the LSO decreases for the binaries modeled by the two codes. At the $1:20$ mass ratio, the time to complete the orbital evolution is identical. This indicates that the two programs yield the same results for well-chosen initial parameters. We have chosen the first one-third of the orbital evolution as the time interval of our interest, due to the large number of simulations required.

\begin{figure}[!h]
			\begin{center}
				\includegraphics[width = 0.5 \textwidth]{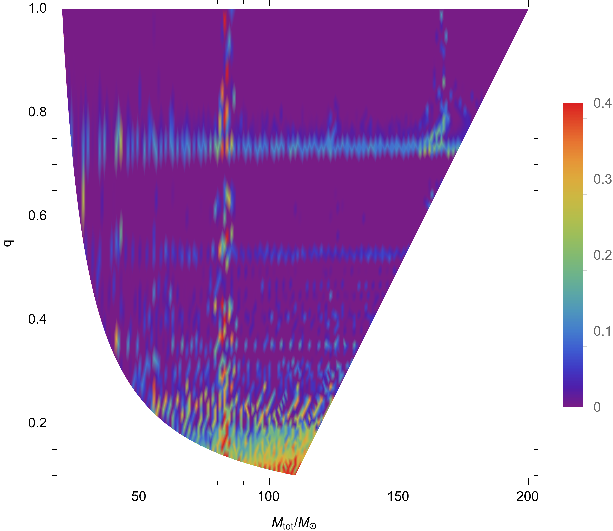}
				\caption{The figure displays the mismatch -- for the non-spinning case -- between the corresponding \texttt{SEOBNRE} and \texttt{CBwaves} waveform over the investigated region of the investigated parameter space. The mass ratio $q$ and the total gravitational mass $M$ are represented along the two axes, while the mismatch is shown on the color bar. The boundaries of the color map are explained by two factors: the straight oblique line on the right is the result of the symmetry of our sampling in $m_1$ and $m_2$, while the boundary on the left is defined by the relationship $10(1 + 1/q)$. Although it largely meets our prior expectations, it still contains some irregularities, which are examined in greater detail below.} \label{fig:mismatch-non-spinning}
			\end{center}
		\end{figure}

		To calculate the mismatch, we used the \texttt{kuibit} \cite{Bozzola2021} python package, which was developed to analyze waveform data generated by \texttt{Einstein Toolkit}. Since the GW data generated by the codes available is identical to what \texttt{kuibit} functions can process, it was an obvious choice for us. Note that the mismatch implemented in the \texttt{kuibit} code is only available for the $(2,2)$ harmonic modes. The algorithm first calculates the overlap between the two GW strains, $h_1$ and $h_2$,
		\begin{equation}
			\mathcal{O} = \frac{\left\langle h_1, h_2 \right\rangle}{\sqrt{\left\langle h_1, h_1 \right\rangle \left\langle h_2, h_2 \right\rangle}},
		\end{equation}
		where
		\begin{equation}
			\left\langle h_1, h_2 \right\rangle = 4\mathfrak{R}\int_{f_\mathrm{max}}^{f_\mathrm{min}} \frac{\tilde{h_1}\tilde{h_2}^*}{S_n(f)} \mathrm{d}f
		\end{equation}
		is the inner product of the two strains, the tildes indicate the Fourier transform, and $S_n(f)$ is the power spectral density. For detectors, $S_n$ is the spectral noise density. Then, the mismatch (or unfaithfulness) is the marginalized overlap over some quantities, which can be given as
		\begin{equation}
			\mathcal{M} = \max \limits_{t,\phi,\psi} \mathcal{O} (h_1, h_2),
		\end{equation}
		where the $\max$ was taken over timeshifts, polarization angles, and phase. Using the above expression of the mismatch, we calculated it at each point with \texttt{kuibit}, shown in Fig. \ref{fig:mismatch-non-spinning}.

\begin{figure}[!h]
			\begin{center}
				\begin{subfigure}{0.48\textwidth}
					\includegraphics[width = \textwidth]{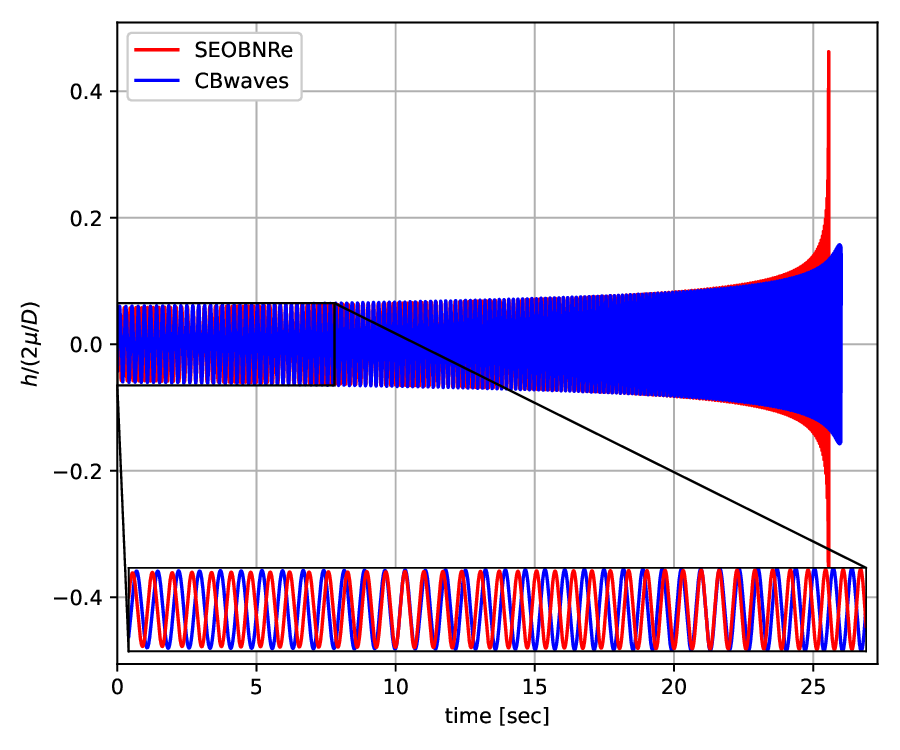}
				\end{subfigure}
				\hfill
				\begin{subfigure}[b]{0.48\textwidth}
					\includegraphics[width = \textwidth]{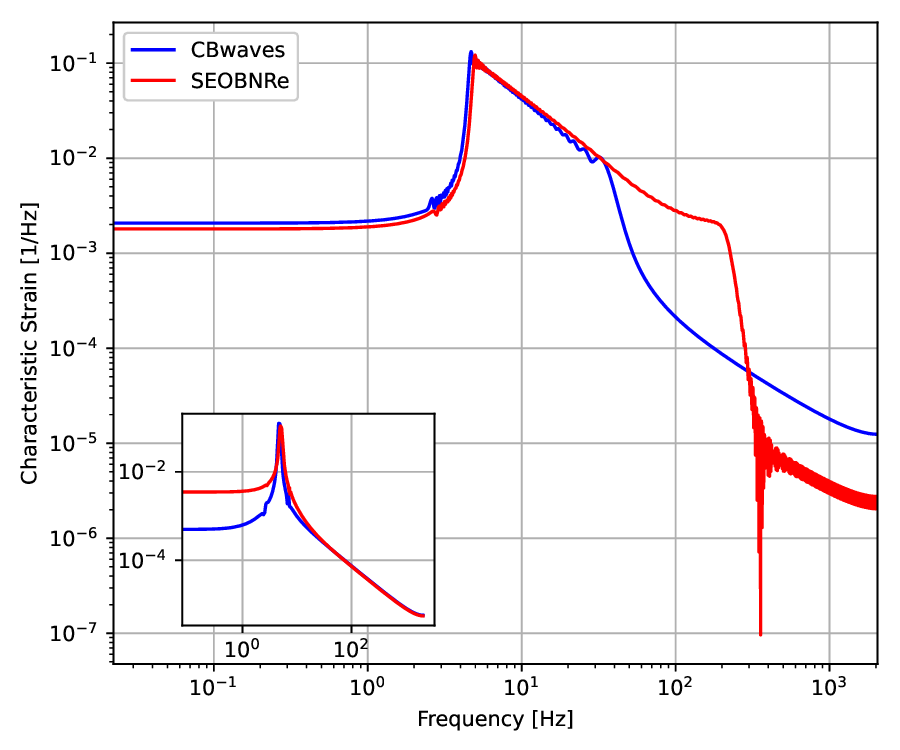}
				\end{subfigure}
				\caption{The right panel of the figure shows the time series of the gravitational wave strain, while the left panel is the characteristic strain, derived from the strain, of the $q \simeq 0.8775$ and $M \simeq 83.6364$ outlier point in Fig. \ref{fig:mismatch-non-spinning}. We magnified on both panels the time window where the mismatch was calculated.} \label{fig:strain-non-spinning-irreg}
			\end{center}
		\end{figure}

		In figure \ref{fig:mismatch-non-spinning}, the boundaries appearing on it can be accounted for by how the mapping of the masses was chosen. On the right side of the map, the straight slanted line occurs due to the symmetry of our configurations in $m_1$ and $m_2$, whereas on the left-hand side, the boundary is delineated by the relationship $10(1 + 1/q)$ relationship. What we can see is partly in line with our prior expectations from Fig. \ref{fig:separation-diffs-s1-06}, however, along the $M \simeq 83.6364$ vertical line we see points where the mismatch is larger than $0.4$.  We decided to take a closer look at the irregularities seen in the figure, and we chose to analyze the GW strain generated by the two codes where $q \simeq 0.8775$ and $M \simeq 83.6364$. In Fig. \ref{fig:strain-non-spinning-irreg}, we show the waveform and characteristic strain of said configuration. We strongly suspect that the two waveforms are out of phase, which is confirmed by Fig. \ref{fig:strain-non-spinning-irreg}. However, we can identify periods when the two are in phase. Then, they move out of phase again. This is a consequence of the differences between the formalisms used in the two software. 
		
\begin{figure}[!h]
			\begin{center}
				\includegraphics[width = 0.7 \textwidth]{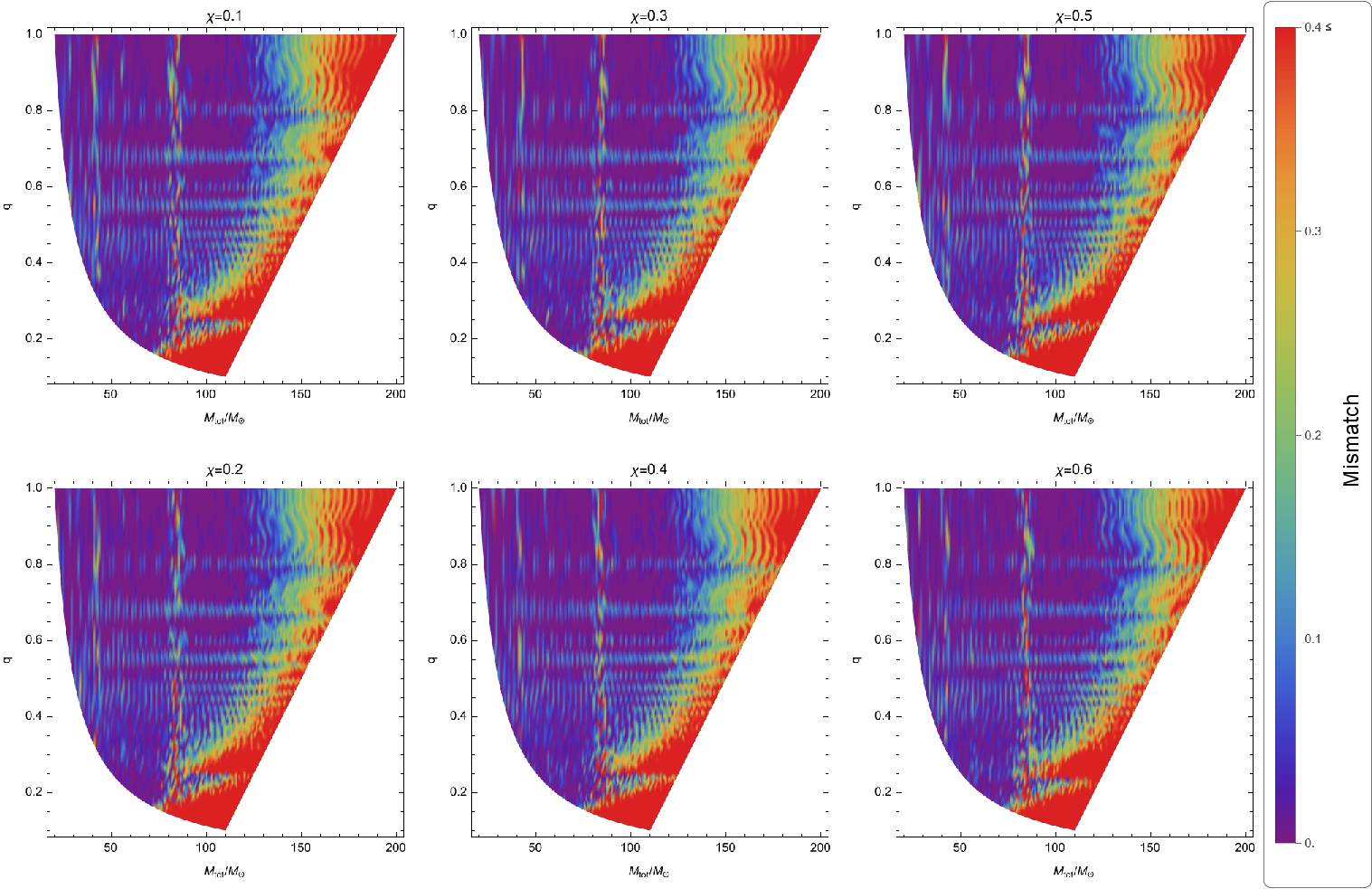}
				\caption{This panel of figures, as in Fig. \ref{fig:mismatch-non-spinning}, shows the mismatches of the two corresponding waveforms for spin-aligned configurations. Here, we collected configurations between the values $[0.1, 0.6]$ of the dimensionless spin parameter.} \label{fig:mismatch-grid-aligned}
			\end{center}
		\end{figure}

\begin{figure}[!h]
			\begin{center}
				\includegraphics[width = 0.7 \textwidth]{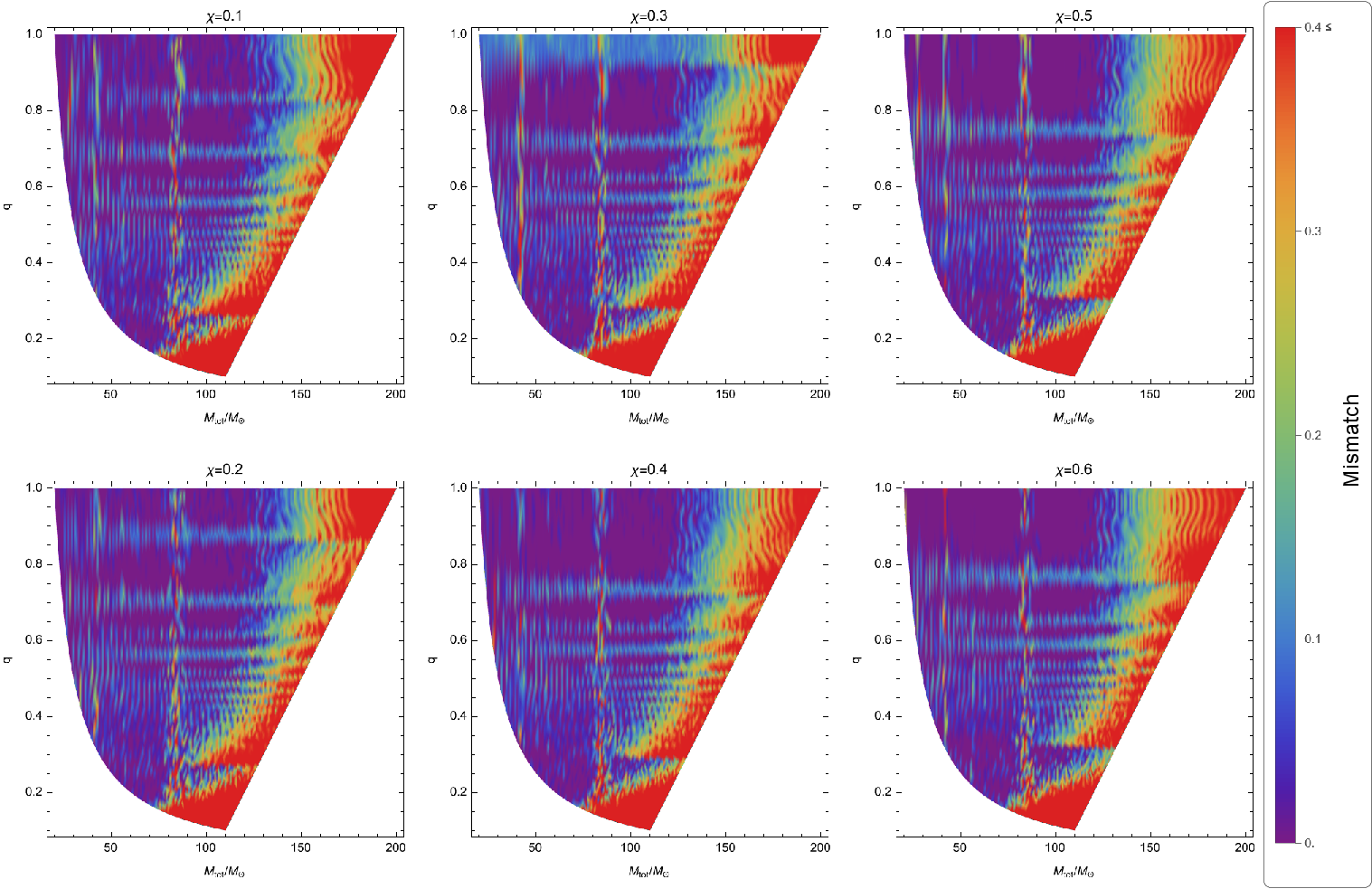}
				\caption{This panel of figures collects the mismatches for the same spin range as in Fig. \ref{fig:mismatch-grid-aligned}, except the two spins were not aligned during the generation of the waveforms in the case of \texttt{CBwaves}.} \label{fig:mismatch-grid-non-aligned}
			\end{center}
		\end{figure}

		In the EOB Hamiltonian, e.g., a spin self-coupling term is introduced. Moreover, the radiation field contains the tail term in the \texttt{SEOBNRE}. It also mentioned in Eq. \ref{eq:eom-cbw-used}, that we used radiation-reaction and spin-spin interaction terms during the computations with \texttt{CBwaves}. We examined in great detail in Fig. \ref{fig:mismatch-grid-aligned} and \ref{fig:mismatch-grid-non-aligned} how spin affects the mismatch map outlined in Fig. \ref{fig:mismatch-non-spinning}. First, let's focus on Fig. \ref{fig:mismatch-grid-aligned}, where we showed the mismatch when the initial spin of the primary component was perpendicular to the orbital plane, i.e. spin-aligned. Here, we immediately see significant differences from Fig. \ref{fig:mismatch-non-spinning}, the red band along the right side of the map. As the spin grows, this band seems to remain relatively unchanged. Also, a second band, associated with high-mismatch values, is present below $50~\mathrm{M}$. The magnitude of the dimensionless spin parameter affects the exact position and prominence of the second band. We also have made calculations where the spin in \texttt{CBwaves} wasn't aligned initially, shown in Fig. \ref{fig:mismatch-grid-non-aligned}. There are significantly more points where the mismatch, even at lower mass ratios, becomes worse than in Fig. \ref{fig:mismatch-grid-aligned}, as expected. At $\chi_1 = 03$, even a band of bit higher mismatch appears between $1$ and $\sim 0.9$ than on any other panel. However, when the dimensionless spin parameter is higher (like $\chi = 0.5$ or $0.6$), we notice larger regions of low mismatches in Fig. \ref{fig:mismatch-grid-non-aligned}. Furthermore, a gap in the band appears between $0.2$ and $0.3$ mass ratios, whose size changes with the spin parameter. Apart from these differences, the panels of Fig. \ref{fig:mismatch-grid-aligned} and \ref{fig:mismatch-grid-non-aligned} are largely similar in appearance, which prompts the question of to what extent the spin of the primary black hole remains aligned throughout its orbital evolution, determined by \texttt{CBwaves}.  To answer that question, we have plotted the orbital evolution of the $q = 1$ mass-ratio configuration for both aligned and non-aligned primary spin, using $\chi = 0.6$ initial value. We have shown it in Fig. \ref{fig:cbw-orbits}, see below.

\begin{figure}[!h]
			\begin{center}
				\begin{subfigure}[b]{0.3\textwidth}
					\caption*{\tiny $\chi_1 = 0$, $q = 1$, $M = 20~\mathrm{M}_\odot$}
					\includegraphics[width = \textwidth]{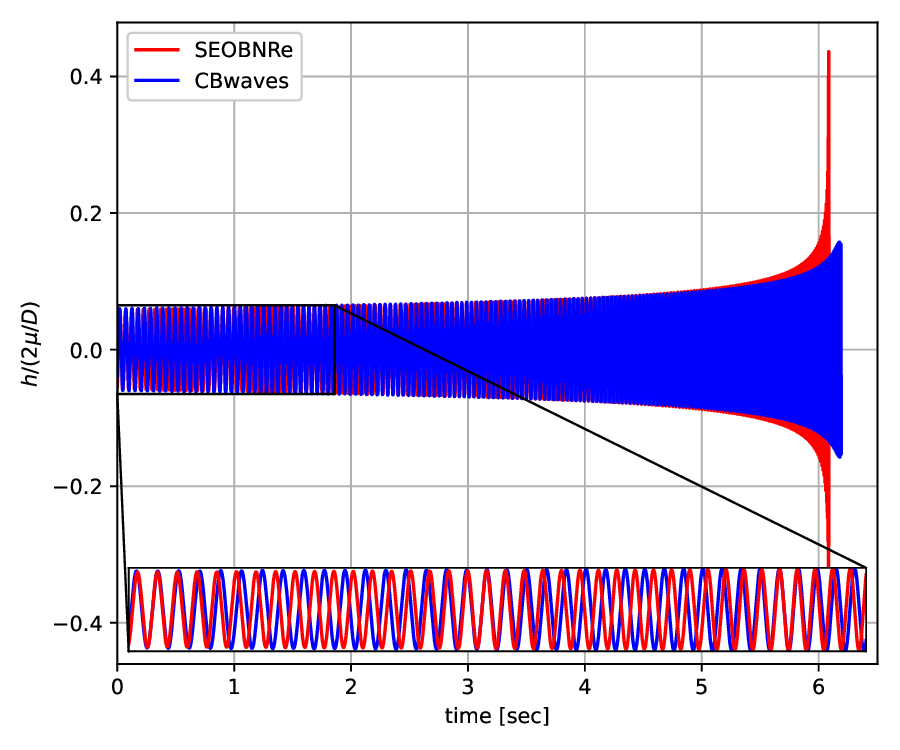}
				\end{subfigure}
				\hfill
				\begin{subfigure}[b]{0.3\textwidth}
					\caption*{\tiny $\chi_1 = 0$, $q = 0.1$, $M = 110~\mathrm{M}_\odot$}
					\includegraphics[width = \textwidth]{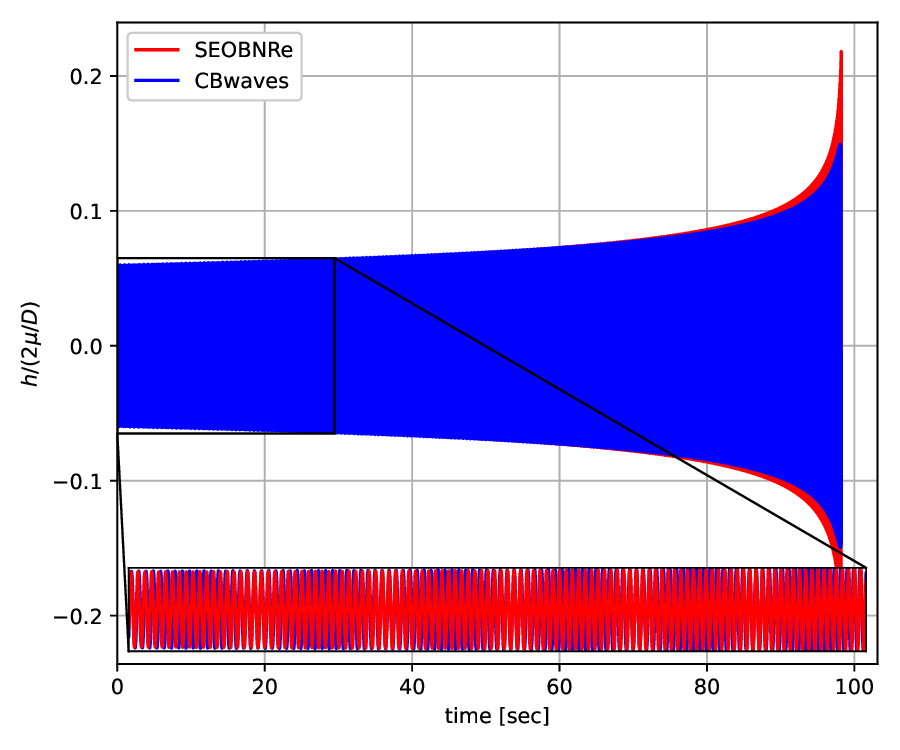}
				\end{subfigure}
				\hfill
				\begin{subfigure}[b]{0.3\textwidth}
					\caption*{\tiny SA,\:$\chi_1\!=\!0.6$,\:$q\!=\!1$,\:$M\!=\!20\:\mathrm{M}_\odot$}
					\includegraphics[width = \textwidth]{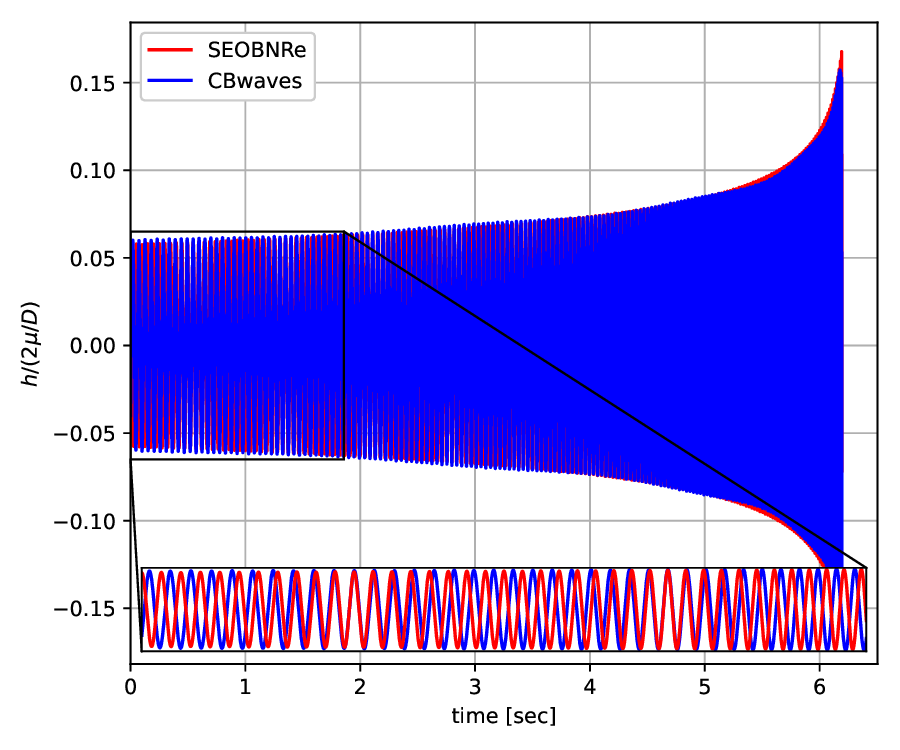}
				\end{subfigure}
				\\
				\begin{subfigure}[b]{0.3\textwidth}
					\caption*{\tiny SA,\:$\chi_1\!=\!0.6$,\:$q\!=\!0.1$,\:$M\!=\!110\:\mathrm{M}_\odot$}
					\includegraphics[width = \textwidth]{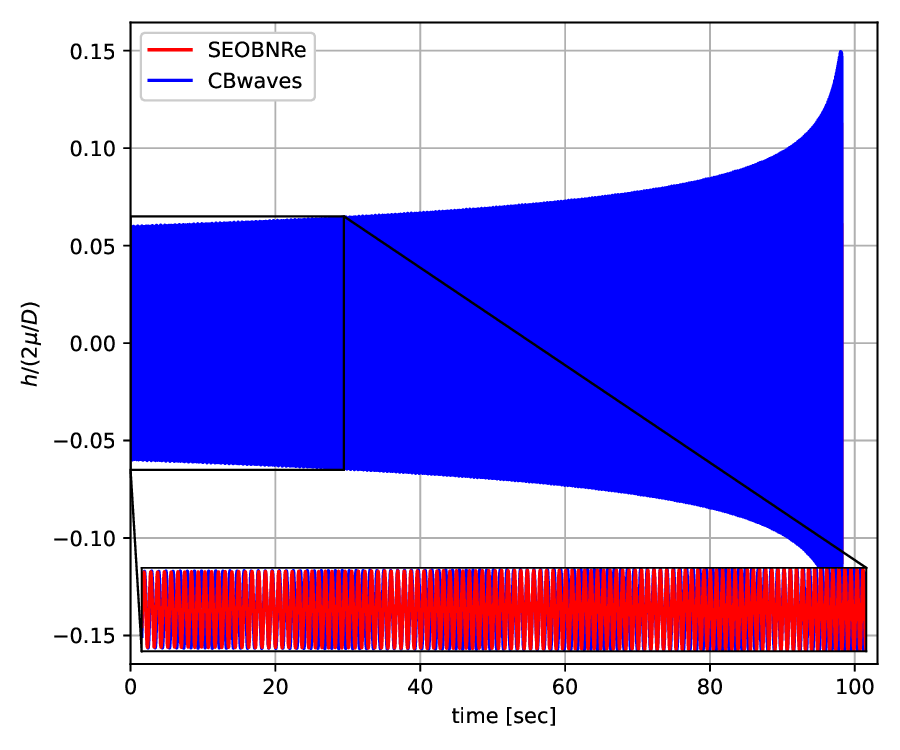}
				\end{subfigure}
				\hfill
				\begin{subfigure}[b]{0.3\textwidth}
					\caption*{\tiny NA,\:$\chi_1\!=\!0.6$,\:$q\!=\!1$,\:$M\!=\!20\:\mathrm{M}_\odot$}
					\includegraphics[width = \textwidth]{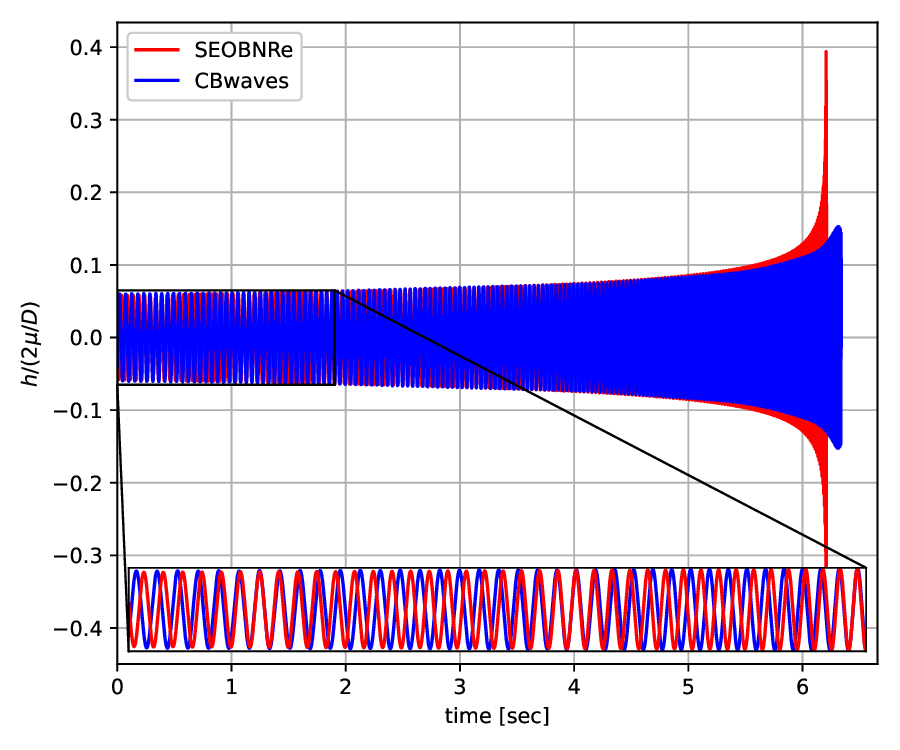}
				\end{subfigure}
				\hfill
				\begin{subfigure}[b]{0.3\textwidth}
					\caption*{\tiny NA,\:$\chi_1\!=\!0.6$,\:$q\!=\!0.1$,\:$M\!=\!110\:\mathrm{M}_\odot$}
					\includegraphics[width = \textwidth]{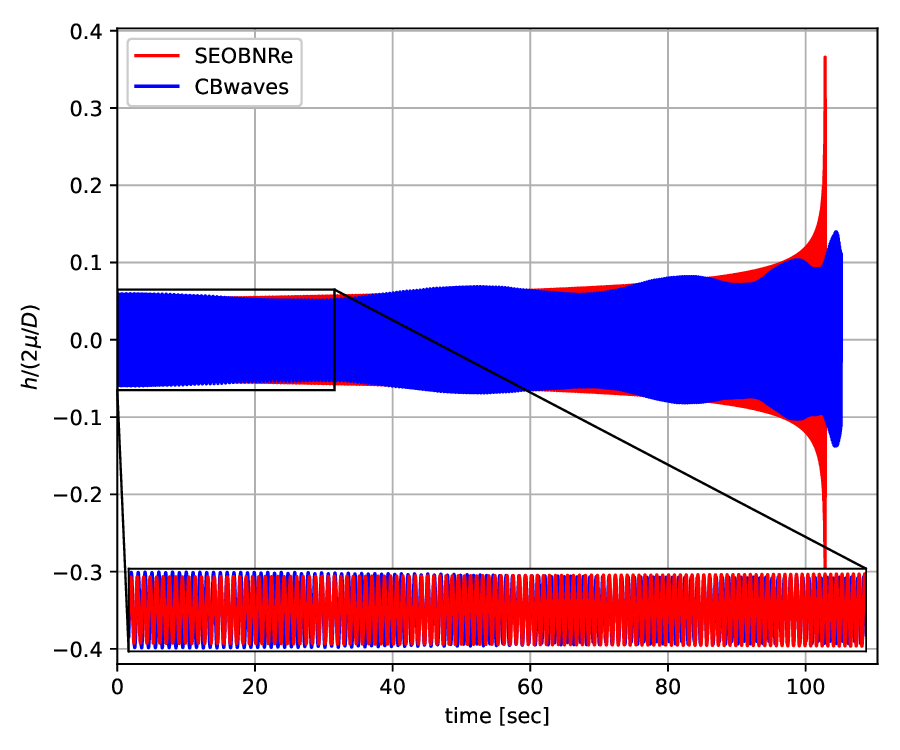}
				\end{subfigure}
				\caption{This panel of figures shows the waveform of selected configurations. Here, we also magnified the part of the strain where the mismatch was calculated.} \label{fig:selected-waveforms}
			\end{center}
		\end{figure}

		On a close examination of Figure \ref{fig:cbw-orbits}, it can be clearly seen that the spin does not remain perpendicular to the orbital plane. In fact, the orbital precession would not be visible on the left panel had the spin retained its alignment. Even though, there are some differences between the two panels, such as the inclination of the orbits and the time spent near the LSO. On the other hand, in Fig. \ref{fig:selected-waveforms}, the spin effects are not apparent in the spin-aligned waveforms, whereas the characteristic spin-induced change is clearly visible in the panel representing the non-aligned $1:10$ mass ratio scenario.

		\begin{figure}[!h]
			\begin{center}
				\begin{subfigure}{0.7 \textwidth}
					\includegraphics[width = \textwidth]{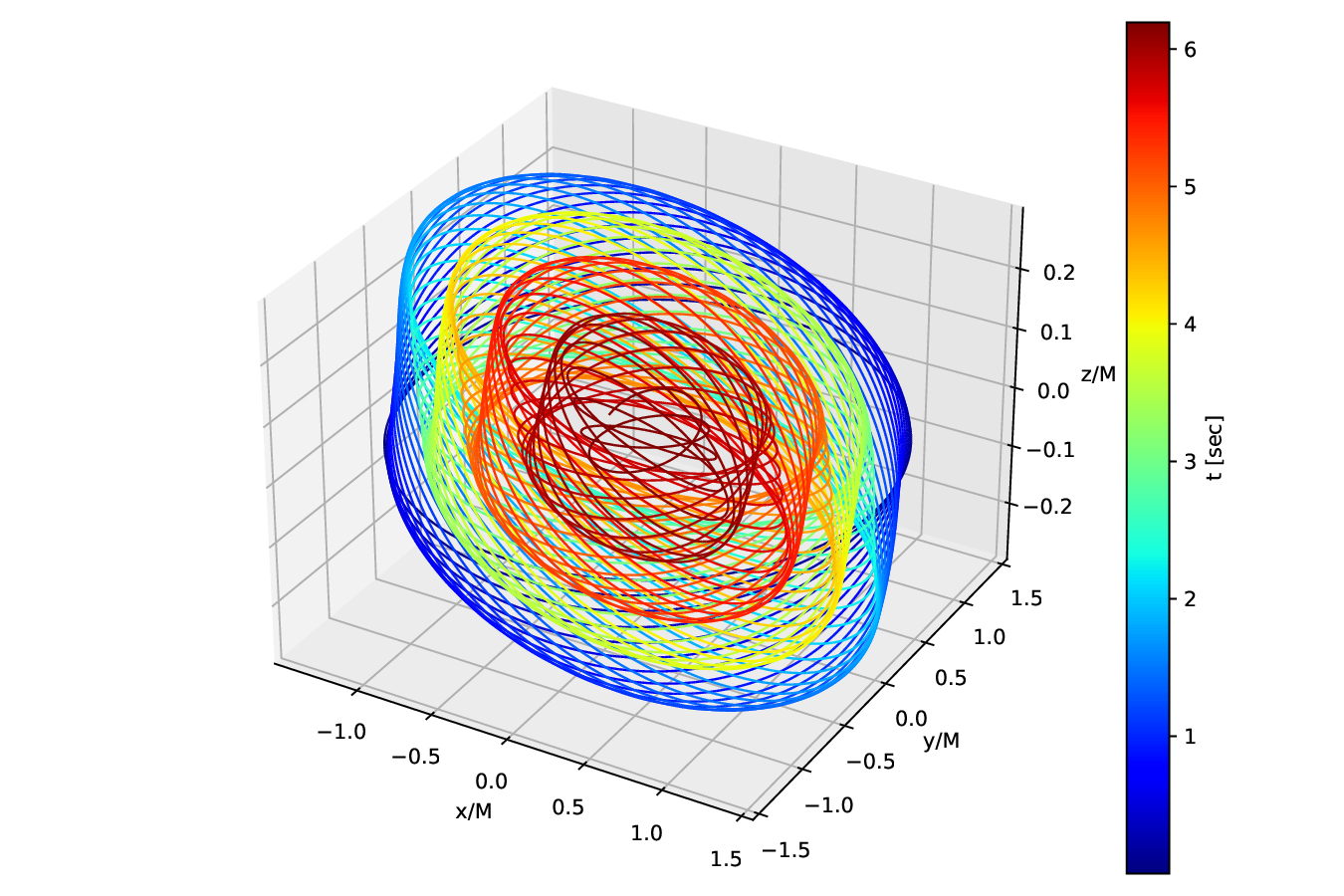}
				\end{subfigure}
				\\
				\begin{subfigure}{0.7 \textwidth}
					\includegraphics[width = \textwidth]{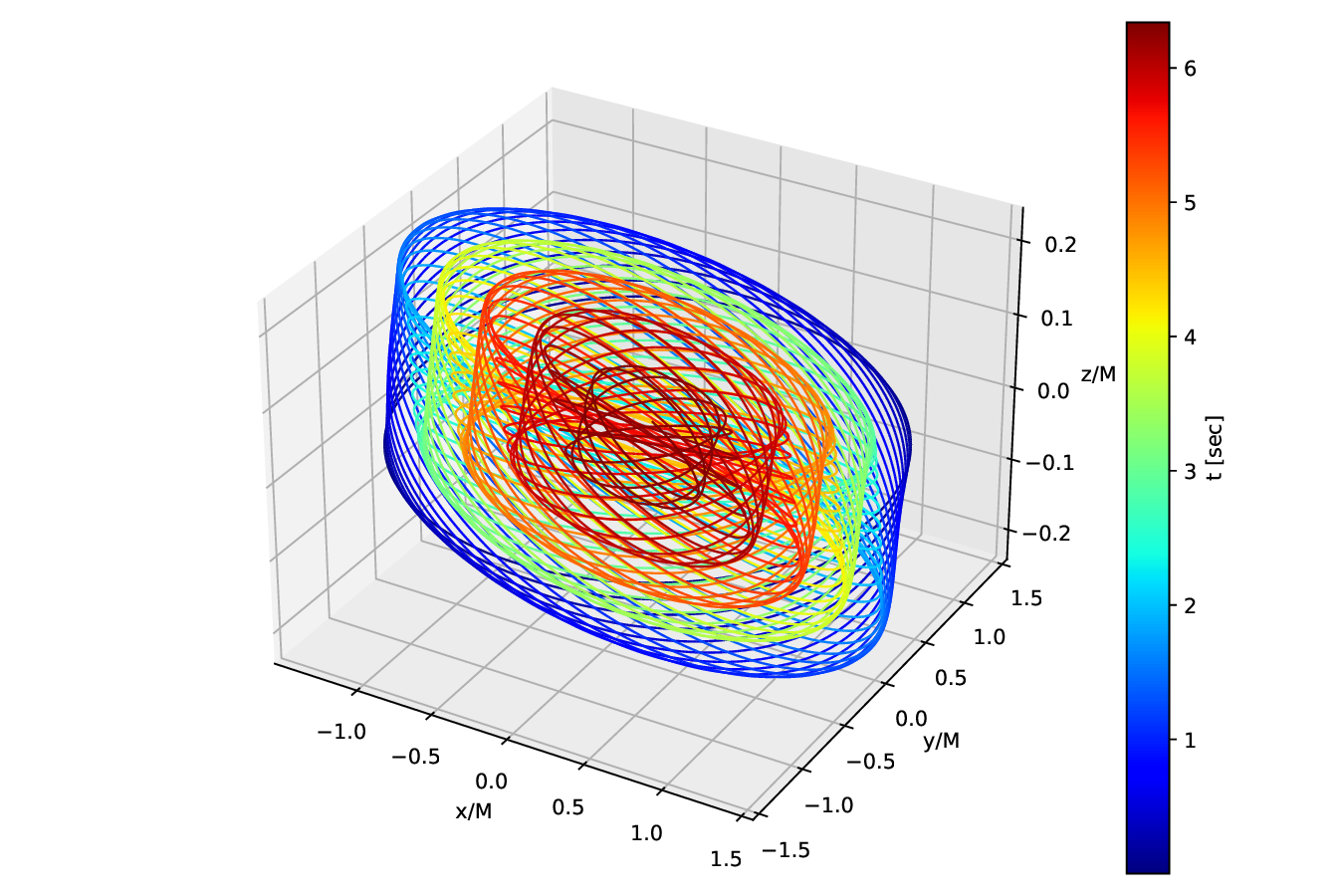}
				\end{subfigure}
				\caption{The orbital evolution of the secondary component is shown in this figure. We plotted the spin-aligned simulation's trajectory, on the top panel, and the trajectory with not-aligned spin is shown on the bottom panel. This proves that the spin does not remain aligned during the orbital evolution in \texttt{CBwaves}. However, small differences can be seen between the two panels, amongst them the most visible is the number of orbits spent close to the LSO.} \label{fig:cbw-orbits}
			\end{center}
		\end{figure}

\begin{figure}[!h]
			\begin{center}
				\begin{subfigure}[b]{0.3\textwidth}
					\caption*{\tiny $\chi_1 = 0$, $q = 1$, $M = 20~\mathrm{M}_\odot$}
					\includegraphics[width = \textwidth]{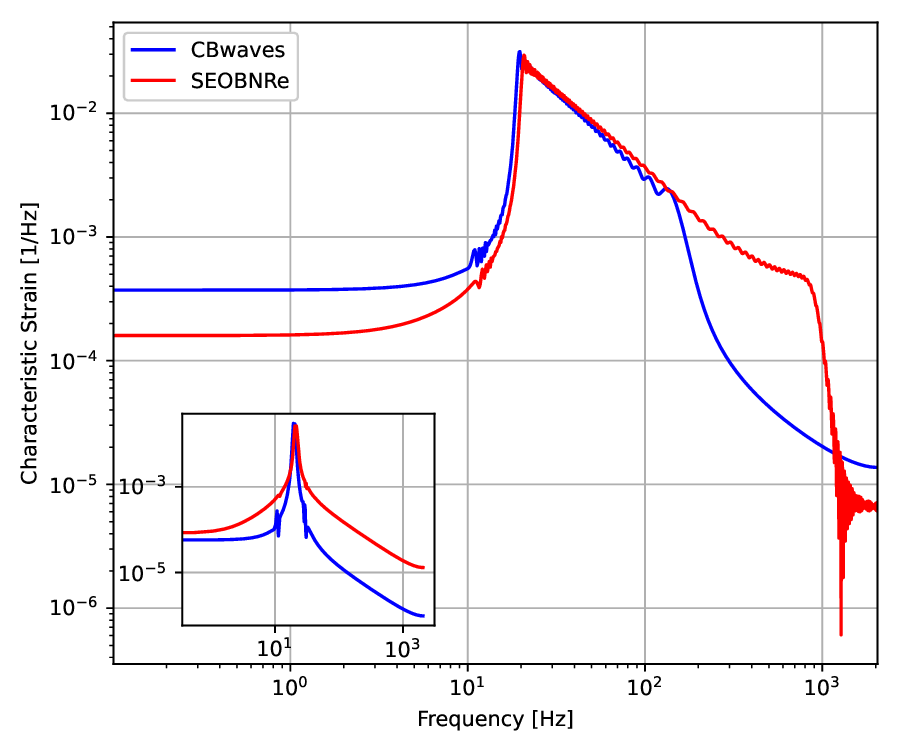}
				\end{subfigure}
				\hfill
				\begin{subfigure}[b]{0.3\textwidth}
					\caption*{\tiny $\chi_1 = 0$, $q = 0.1$, $M = 110~\mathrm{M}_\odot$}
					\includegraphics[width = \textwidth]{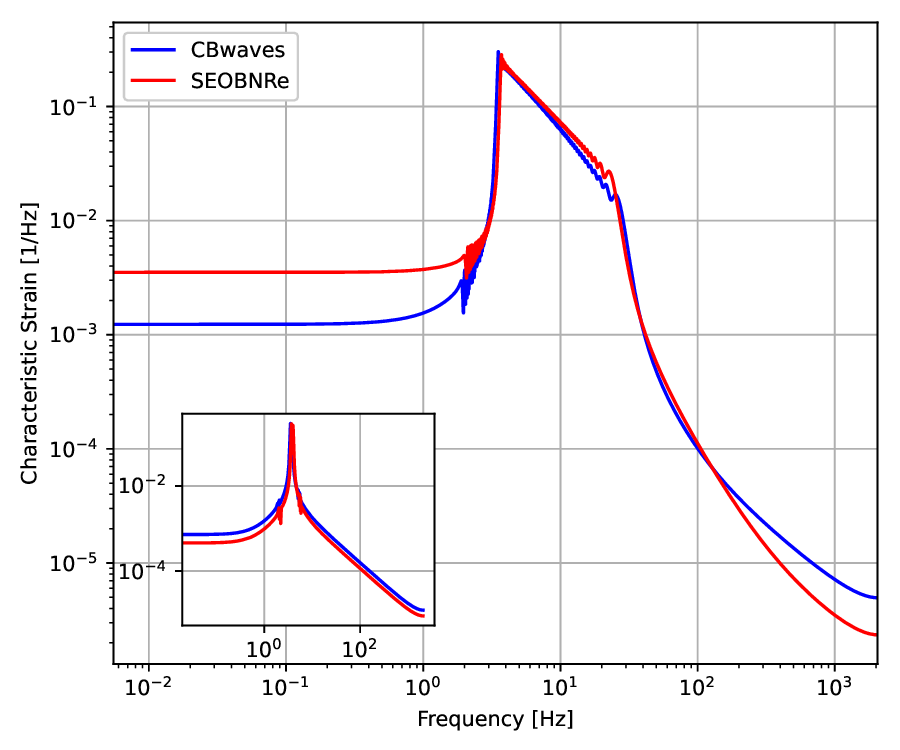}
				\end{subfigure}
				\hfill
				\begin{subfigure}[b]{0.3\textwidth}
					\caption*{\tiny SA,\:$\chi_1\!=\!0.6$,\:$q\!=\!1$,\:$M\!=\!20\:\mathrm{M}_\odot$}
					\includegraphics[width = \textwidth]{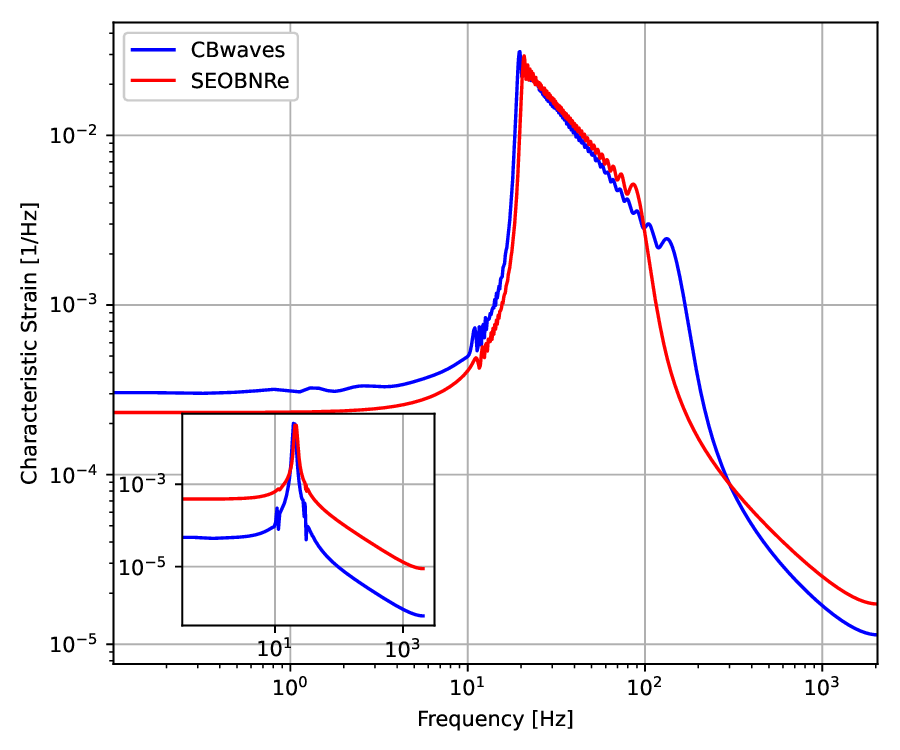}
				\end{subfigure}
				\\
				\begin{subfigure}[b]{0.3\textwidth}
					\caption*{\tiny SA,\:$\chi_1\!=\!0.6$,\:$q\!=\!0.1$,\:$M\!=\!110\:\mathrm{M}_\odot$}
					\includegraphics[width = \textwidth]{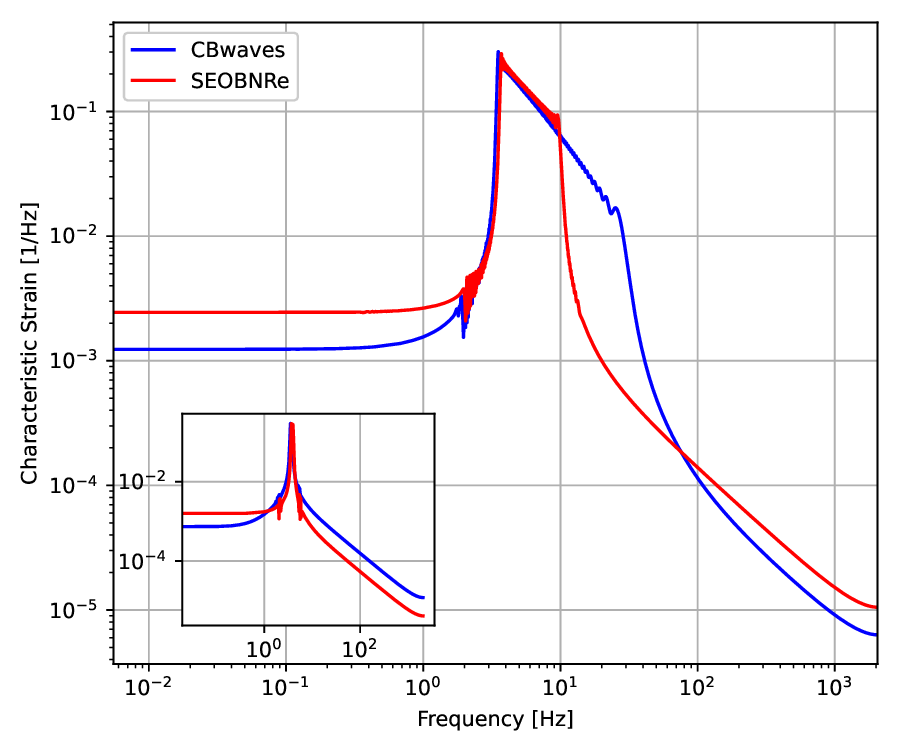}
				\end{subfigure}
				\hfill
				\begin{subfigure}[b]{0.3\textwidth}
					\caption*{\tiny NA,\:$\chi_1\!=\!0.6$,\:$q\!=\!1$,\!$M\!=\!20\:\mathrm{M}_\odot$}
					\includegraphics[width = \textwidth]{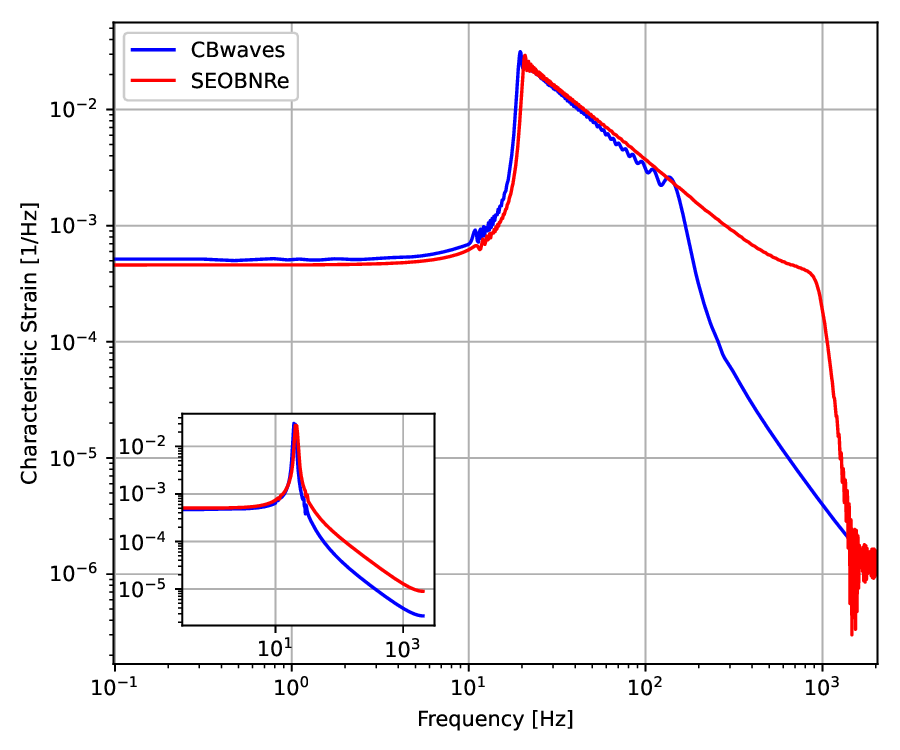}
				\end{subfigure}
				\hfill
				\begin{subfigure}[b]{0.3\textwidth}
					\caption*{\tiny NA,\:$\chi_1\!=\!0.6$,\:$q\!=\!0.1$,\:$M\!=\!110\:\mathrm{M}_\odot$}
					\includegraphics[width = \textwidth]{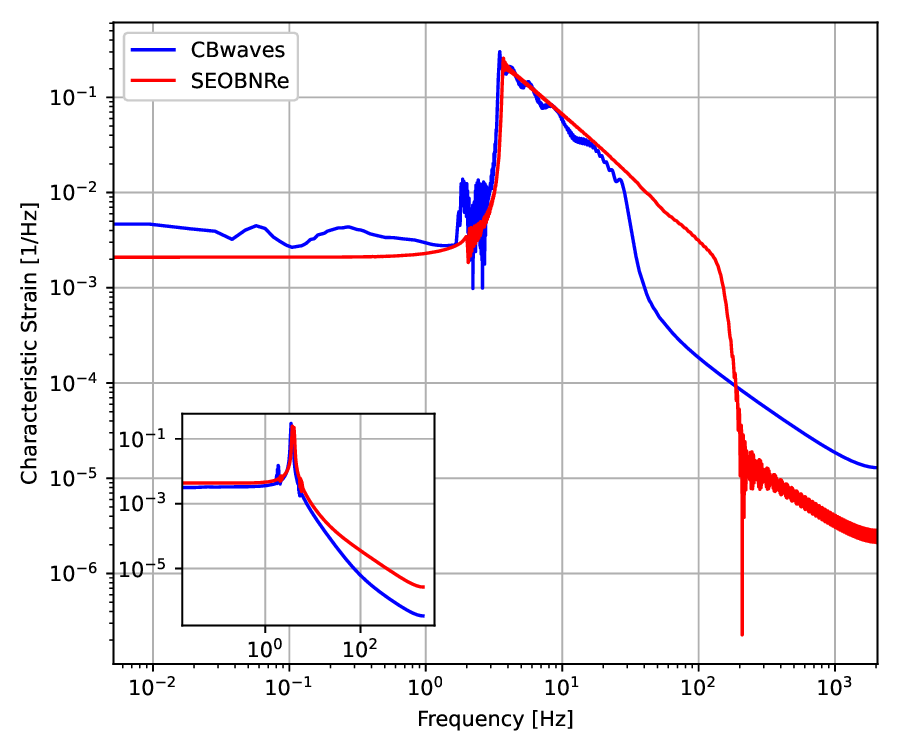}
				\end{subfigure}
				\caption{In this panel, we collected the characteristic strains of the same selected configurations as in Fig. \ref{fig:selected-waveforms}. Note that, the characteristic strain shown in the magnified panels is derived from the waveforms in our chosen time window.} \label{fig:selected-characteristic-strain}
			\end{center}
		\end{figure}

		In Fig. \ref{fig:selected-waveforms}, we collected the waveform of two configurations from the not-spinning, $\chi_1 = 0.6$ spin-aligned and not-aligned mismatch panels. We selected the points with $(q = 1, M = 20~\mathrm{M}_\odot)$ and $(q = 0.1, M = 110~\mathrm{M}_\odot)$, the two extremes on the map. As mentioned above, even though we chose our initial spin to be perpendicular to the orbit it doesn't remain as. Because of that the orbits calculated by \texttt{CBwaves} are precessing, however, it doesn't show on the waveform. We can see how the spin affects the waveform in the bottom 3rd panel of Fig. \ref{fig:selected-waveforms}. In Fig. \ref{fig:selected-characteristic-strain}, we show the characteristic strain of the waveforms shown in Fig. \ref{fig:selected-waveforms}. The characteristic strain can be obtained by Fourier transforming the amplitude of the strain,
		\begin{equation}
				\tilde{h} = \mathcal{F}\{h(t)\}(f) = \int \limits_{-\infty}^\infty h(t)e^{-2\pi ift}~dt,
		\end{equation}
		as it can be found in \cite{MCB15}.

\newpage

	\section{Conclusion}
First by looking at Fig. \ref{fig:separation-diffs-s1-06}, we see that at $1:20$ mass-ratio the separation computed by both codes is in close agreement. For configurations with $q < 1/20$ the separation of $6~\mathrm{M}$ where the simulation was terminated is reached earlier by \texttt{SEOBNRE}, in contrast to that, when $q > 1/20$ the same limit was reached earlier by \texttt{CBwaves}. The larger the difference in mass-ratio from $q \simeq 1/20$, the greater the difference in the evolution time.   We made detailed contour maps for the mismatch (or unfaithfulness) of various spin configurations, e.g. aligned and non-aligned spins, see Fig. \ref{fig:mismatch-non-spinning}, \ref{fig:mismatch-grid-aligned}, and \ref{fig:mismatch-grid-non-aligned}. We found that as the mass-ratio is close to $1:10$, the mismatch between the two models grows larger. A similar behavior is exhibited toward larger total masses with spins, but irrespective of the alignment. Bends of outlier mismatches appear on the maps at certain values of total masses and mass-ratios, the contrast and number of the bends indicate a dependence on the spin magnitude. By comparing Fig. \ref{fig:mismatch-grid-aligned} and \ref{fig:mismatch-grid-non-aligned}, we concluded that the spins did not retain the initial alignment set in \texttt{CBwaves}, as shown in Fig. \ref{fig:cbw-orbits}. However, the effects of the spin, in Fig. \ref{fig:selected-waveforms}, are unnoticeable on the aligned waveforms. The major advantages of \texttt{CBwaves}, summarized in \cite{CBwaves}, over \texttt{SEOBNRE} lies in larger freedom in the choice of initial parameter values due to the underlying post-Newtonian extension and design choices in the implementation of it. In accordance with observational results, the code has proven its accuracy in modeling \cite{Barta2018, KaVas22} compact binary inspiral with extreme astrophysical parameters, such as extreme mass-ratio, supermassive components, high eccentricity, and spin.

	\section{Acknowledgement}
First and foremost, we are immensely grateful to the Wigner Scientific Computing Laboratory for providing us with the necessary computational infrastructure. Without it, this work wouldn't have been prepared in such a short time. We wish to express our gratitude for the advice our colleague, Gergely \'A  L\'aszl\'o provided. This research has been supported in part by the National Research, Development and Innovation Office (Hungarian abbreviation: NKFIH) under OTKA grant agreement No. K138277. The authors present this work as a tribute to their late mentor and colleague, M\'aty\'as Zsolt Vas\'uth, who passed away earlier this year.

\bibliographystyle{unsrturl}
\bibliography{seobnre.bib}

\end{document}